\documentclass[aps,twocolumn,pra,superscriptaddress,showpacs,tightenlines]{revtex4}
\usepackage{amssymb}
\usepackage{amsmath}
\usepackage{graphicx}
\usepackage{epsfig}
\usepackage{subfigure}
\usepackage{amsfonts}
\usepackage{txfonts}

\begin{document}
\title{Quantum thermalization of two coupled two-level systems
in eigenstate and bare-state representations}
\author{Jie-Qiao Liao}
\affiliation{Institute of Theoretical Physics, Chinese Academy of
Sciences, Beijing 100190, China}
\affiliation{Department of Physics
and Institute of Theoretical Physics, The Chinese University of Hong
Kong, Shatin, Hong Kong Special Administrative Region, China}
\author{Jin-Feng Huang}
\affiliation{Key Laboratory of Low-Dimensional Quantum Structures
and Quantum Control of Ministry of Education, and Department of
Physics, Hunan Normal University, Changsha 410081, China}
\affiliation{Institute of Theoretical Physics, Chinese Academy of
Sciences, Beijing 100190, China}
\author{Le-Man Kuang}
\affiliation{Key Laboratory of Low-Dimensional Quantum Structures
and Quantum Control of Ministry of Education, and Department of
Physics, Hunan Normal University, Changsha 410081, China}

\date{\today}
\begin{abstract}
We study analytically the quantum thermalization of two coupled
two-level systems (TLSs), which are connected with either two
independent heat baths (IHBs) or a common heat bath (CHB). We
understand the quantum thermalization in eigenstate and bare-state
representations when the coupling between the two TLSs is stronger
and weaker than the TLS-bath couplings, respectively. In the IHB
case, we find that when the two IHBs have the same temperatures, the
two coupled TLSs in eigenstate representation can be thermalized
with the same temperature as those of the IHBs. However, in the case
of two IHBs at different temperatures, just when the energy detuning
between the two TLSs satisfies a special condition, the two coupled
TLSs in eigenstate representation can be thermalized with an
immediate temperature between those of the two IHBs. In bare-state
representation, we find a counterintuitive phenomenon that, under
some conditions, the temperature of the TLS connected with the
high-temperature bath is lower than that of the other TLS, which is
connected with the low-temperature bath. In the CHB case, the
coupled TLSs in eigenstate representation can be thermalized with
the same temperature as that of the CHB in nonresonant cases. In
bare-state representation, the TLS with a larger energy separation
can be thermalized to a thermal equilibrium with a lower
temperature. In the resonant case, we find a phenomenon of
anti-thermalization. We also study the steady-state entanglement
between the two TLSs in both the IHB and CHB cases.
\end{abstract}
\narrowtext



\maketitle \narrowtext

\section{introduction}

Conventional quantum thermalization~\cite{Breuer,Gemmerbook} is
understood as an irreversibly dynamic process under which a quantum
system immersed in a heat bath approaches a thermal equilibrium
state with the same temperature as that of the bath. The thermal
equilibrium state~\cite{Greiner} of the thermalized system reads as
$\rho_{th}(T)=Z_{S}^{-1}\exp[-H_{S}/(k_{B}T)]$, which is merely
determined by the Hamiltonian $H_{S}$ of the thermalized system and
the temperature $T$ of its environment, where
$Z_{S}=\text{Tr}_{S}\{\exp[-H_{S}/(k_{B}T)]\}$ is the partition
function of the thermalized system, with $k_{B}$ being the Bolztmann
constant. During the course of a quantum thermalization, all of the
initial information of the thermalized system is totally erased by
its environment. Recently, much attention has been paid to quantum
thermalization (e.g.,
Refs.~\cite{Rigolnature,Rigol,Berman,Reimann,Linden,Rajagopal,Tasaki,Liao,Lychkovskiy}).
Specifically, a new kind of thermalization, called canonical
thermalization (e.g.,
Refs.~\cite{Popescu,Goldstein,Gemmer,Dong,Reimann2010}), has been
proposed.

The conventional quantum thermalization works for the situations
wherein one quantum system is connected with just one environment at
thermal equilibrium. When we consider a composite quantum system,
which is constructed with many subsystems and connected with many
environments, the conventional quantum thermalization is no longer
valid. Therefore, the density operator of the composite quantum
system at thermal equilibrium can not be written as $\rho_{th}(T)$.
In fact, quantum thermalization of a composite quantum system is a
very complex problem. On one hand, from the viewpoint of the
environments, the composite quantum system could be connected with
many independent environments or a common environment. At the same
time, the temperatures of these environments could be the same or
different. On the other hand, the coupling strengths among these
subsystems can affect the physical picture to describe the quantum
thermalization of the composite quantum system. When the coupling
strengths among these subsystems are stronger than the system-bath
couplings, the composite quantum system can be regarded as a single
system, while it is regarded as many individual subsystems when the
couplings among them are weaker than the system-bath couplings.

The above mentioned situations come true in recent years since
quantum systems can be manufactured to be more and more complicated
and small, based on the great advances in physics, chemistry, and
biology~\cite{Bustamante}. Therefore, the research on
\textit{thermodynamics of small systems} becomes very interesting.
In particular, quantum thermalization of composite quantum systems
becomes an important topic since many important results in this
field, such as nonequilibrium work relations and fluctuation theorem
(e.g.,
Refs.~\cite{Jarzynski,Crooks,Tasaki2000,Evans2002,Hanggi2009}), are
based on the thermal equilibrium state of the composite quantum
systems. As composite quantum systems are composed of many
subsystems, they could be connected either with many independent
environments at different temperatures or a common environment.
Therefore, it is natural to ask the following questions: (i) How do
the couplings among the subsystems affect the quantum thermalization
of a composite quantum system? (ii) What are the steady-state
properties of a composite quantum system when it is connected with
either many independent environments or a common environment?

With these questions, in this paper, we study the quantum
thermalization of two coupled two-level systems (TLSs) that are
immersed in either two independent heat baths (IHBs) or a common
heat bath (CHB). Simple as this model is, it is illustrative. When
the coupling between the two TLSs is stronger than the TLS-bath
couplings, the two TLSs can be considered as an effective composite
system, i.e., a four-level system, and then we understand the
quantum thermalization in eigenstate representation of the composite
system. In addition, when the coupling between the two TLSs is
weaker than the TLS-bath couplings, we understand the quantum
thermalization from the viewpoint of each individual TLS. However,
due to the TLS-TLS coupling, the effective temperatures of the two
TLSs should be different from those for the decoupling case.

As for the environments, there are two kinds of different
situations: the IHB case and the CHB case. In the IHB case, we find
that, when the two IHBs have the same temperatures, the two coupled
TLSs in the eigenstate representation can be thermalized with the
same temperature as those of the IHBs. However, in the case where
the two IHBs have different temperatures, just when the energy
detuning between the two TLSs satisfies a special condition, the two
coupled TLSs in eigenstate representation can be thermalized with an
immediate temperature between those of the two IHBs. In the
bare-state representation, we find a counterintuitive phenomenon
that, under some conditions, the temperature of the TLS connected
with the high-temperature heat bath is lower than that of the other
TLS, which is connected with the low-temperature heat bath.

In the CHB case, we also study the quantum thermalization in
eigenstate and bare-state representations. In the eigenstate
representation, the present case reduces to the conventional quantum
thermalization, [i.e., one quantum system (an effective four-level
system formed by the two coupled TLSs) is thermalized by one
environment (the common heat bath) in thermal equilibrium]. In
addition, we also investigate the effective temperatures of the two
TLSs in bare-state representation. It is found that the TLS with a
larger energy separation can be thermalized with a lower
temperature. In particular, in the resonant case, we find a quantum
phenomenon of anti-thermalization when the two TLSs are connected
with a common heat bath.

This paper is organized as follows: In Sec.~\ref{Sec:2}, we present
the physical models and their Hamiltonians. In Sec.~\ref{Sec:3}, we
study the quantum thermalization of the two coupled TLSs immersed in
two IHBs. In Sec.~ \ref{Sec:4}, we consider the case wherein the two
TLSs are immersed in a CHB. We conclude this work in
Sec.~\ref{Sec:5}. Finally, we give two
appendices,~\ref{appeindependentbath} and~\ref{appecommontbath}, for
detailed derivations of quantum master equations and transition
rates for the IHB and CHB cases, respectively.

\section{\label{Sec:2}Physical models and Hamiltonians}

\begin{figure}[tbp]
\includegraphics[bb=33 311 498 776, width=3.3 in]{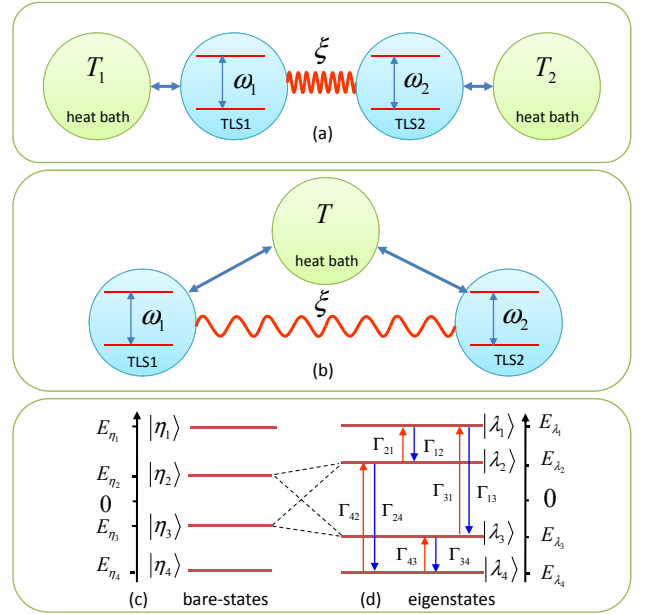}
\caption{(Color online) Schematic diagram of the physical models
under consideration. Two coupled two-level systems, denoted by
TLS$1$ and TLS$2$ with respective energy separations
$\protect\omega_{1}$ and $\protect\omega _{2}$, are connected with
either (a) two IHBs at temperatures $T_{1}$ and $ T_{2}$ or (b) a
CHB at temperature $T$. Between the two TLSs, there exists a
dipole-dipole interaction of strength $\protect\xi$. (c) The levels
of the four bare states $|\protect\eta_{i}\rangle$ $(i=1,2,3,4)$ of
the two free TLSs. (d) The levels of the four eigenstates
$|\protect\lambda_{i}\rangle$ $(i=1,2,3,4)$ of
Hamiltonian~(\protect\ref{HofTLSs}) for the two coupled TLSs. In the
presence of the baths, there exist bath-induced exciting and damping
processes among the four eigenstates. The effective transition rates
from states $|\protect\lambda_{i}\rangle$ to
$|\protect\lambda_{j}\rangle$ are denoted by $\Gamma_{ij}$. Other
cross-dephasing processes are not denoted explicitly.}
\label{schematic}
\end{figure}

Let us start with introducing the physical models [as illustrated in
Figs.~\ref{schematic}(a) and~\ref{schematic}(b)]: two TLSs, denoted
by TLS$1$ and TLS$2$ with respective energy separations $\omega_{1}$
and $\omega_{2}$, couple with each other via a dipole-dipole
interaction of strength $\xi$. At the same time, the two TLSs couple
inevitably with the environments surrounding them. Specifically, in
this paper, we consider two kinds of different cases: the IHB case
and the CHB case. In the former case, the two TLSs are immersed in
two IHBs, while, in the latter case, the two TLSs are immersed in a
CHB.

The Hamiltonian of the total system, including the two TLSs and their
environments, is composed of three parts:
\begin{equation}
H=H_{\text{TLSs}}+H_{B}+H_{I},
\end{equation}
where $H_{\text{TLSs}}$ is the Hamiltonian of the two coupled TLSs, $H_{B}$
is the Hamiltonian of the heat baths, and $H_{I}$ describes the interaction
Hamiltonian between the two TLSs and their baths.

The Hamiltonian $H_{\text{TLSs}}$ (with $\hbar=1$) reads as
\begin{equation}
H_{\text{TLSs}}=\frac{\omega _{1}}{2}\sigma_{1}^{z}+\frac{\omega
_{2}}{2}\sigma_{2}^{z}+\xi(\sigma_{1}^{+}\sigma_{2}^{-}+\sigma_{1}^{-}\sigma
_{2}^{+}).  \label{HofTLSs}
\end{equation}
The first two terms in Eq.~(\ref{HofTLSs}) are free Hamiltonians of
the two TLSs, which are described by the operators $\sigma_{l}^{+}=(
\sigma_{l}^{-})^{\dag}=\vert e\rangle_{ll}\langle g\vert$ and $
\sigma_{l}^{z}=\vert e\rangle_{ll}\langle e\vert-\vert
g\rangle_{ll}\langle g\vert$ $(l=1,2)$, where $|g\rangle_{l}$ and
$|e\rangle_{l}$ are, respectively, the ground and excited states of
the $l$th TLS (i.e., TLS$l$). The last term in Eq.~(\ref{HofTLSs})
describes a dipole-dipole interaction of strength $\xi$ between the
two TLSs. We note that the Hamiltonian given in Eq.~(\ref{HofTLSs})
has been widely studied in various physical problems, such as
quantum logic gates~\cite{Petrosyan}, coherent excitation energy
transfer~\cite{Liao2010}, decoherence
dynamics~\cite{Ferraro2009,Sinaysky2009,Ban2009}, and nonequilibrium
thermal entanglement~\cite{Quiroga2007,Sinaysky2008}.

The Hilbert space of the two TLSs may be spanned by the following
four bare states $|\eta_{1}\rangle=|ee\rangle$,
$|\eta_{2}\rangle=|eg\rangle$, $|\eta_{3}\rangle=|ge\rangle$, and
$|\eta_{4}\rangle=|gg\rangle$ [as shown in Fig.~\ref{schematic}(c)],
which are eigenstates of the free Hamiltonian
$(\omega_{1}\sigma^{z}_{1}+\omega_{2}\sigma^{z}_{2})/2$ of the two
TLSs, with the corresponding eigenenergies
$E_{\eta_{1}}=-E_{\eta_{4}}=\omega_{m}$ and
$E_{\eta_{2}}=-E_{\eta_{3}}=\Delta\omega/2$. Here we have introduced
the mean energy separation $\omega_{m}=(\omega_{1}+\omega_{2})/2$
and the energy detuning $\Delta\omega=\omega_{1}-\omega_{2}$.

Due to the dipole-dipole interaction, a stationary single-excitation
state should be delocalized and composed of a combination of the
single-excitation states in the two TLSs. According to
Hamiltonian~(\ref{HofTLSs}), we can solve the eigenequation
$H_{\text{TLSs}}\vert\lambda
_{n}\rangle=E_{\lambda_{n}}\vert\lambda_{n}\rangle$, $(n=1,2,3,4)$
to obtain the following four eigenstates [as shown in
Fig.~\ref{schematic}(d)]:
\begin{gather}
\vert\lambda_{1}\rangle=|\eta_{1}\rangle,\hspace{0.3 cm}
\vert\lambda _{2}\rangle=\cos(\theta/2)|\eta_{2}\rangle
+\sin(\theta/2)|\eta_{3}\rangle,\nonumber\\
\vert\lambda_{3}\rangle=-\sin(\theta/2)|
\eta_{2}\rangle+\cos(\theta/2)|\eta_{3}\rangle,\hspace{0.3 cm}
\vert\lambda_{4}\rangle=|\eta_{4}\rangle,
\end{gather}
with the corresponding eigenenergies
$E_{\lambda_{1}}=-E_{\lambda_{4}}=\omega_{m}$ and
$E_{\lambda_{2}}=-E_{\lambda_{3}}=\sqrt{\Delta\omega^{2}/4+\xi^{2}}$.
Here we have introduced a mixing angle $\theta$ $(0<\theta<\pi)$ by
$ \tan\theta=2\xi/\Delta\omega$. For a positive $\xi$, when $
\omega_{1}>\omega_{2}$, namely $\Delta\omega>0$, we choose $
\theta=\arctan(2\xi/\Delta\omega)$; When $\omega_{1}<\omega_{2}$,
that is $ \Delta\omega<0$, we choose
$\theta=\pi+\arctan(2\xi/\Delta\omega)$. The dipole-dipole
interaction mixes the two bare states $|\eta_{2}\rangle$ and $
|\eta_{3}\rangle$ with one excitation, and does not change the bare
states $ |\eta_{1}\rangle$ with two excitations and
$|\eta_{4}\rangle$ with zero excitation.

Aside from the dipole-dipole interaction between the two TLSs, there
exist couplings between the TLSs and their environments. In general,
when the couplings of a system with its environment are weak, it is
universal to model the environment as a harmonic-oscillator heat
bath and choose a linear coupling between the system and its
environment~\cite{Leggettnp}. In this paper, we consider this
situation and model the environments as harmonic-oscillator heat
baths. As mentioned above, we will consider two kinds of different
cases: one is the IHB case and the other is the CHB case.

In the IHB case, as shown in Fig.~\ref{schematic}(a), the two TLSs
are immersed in two IHBs described by the Hamiltonian
\begin{gather}
H^{(\text{IHB})}_{B}=H^{(a)}_{B}+H^{(b)}_{B},\nonumber\\
H^{(a)}_{B}=\sum_{j}\omega _{aj}a_{j}^{\dagger }a_{j},\hspace{0.5
cm}H^{(b)}_{B}=\sum_{k}\omega _{bk}b_{k}^{\dagger }b_{k}.
\end{gather}
Here $H^{(a)}_{B}$ and $H^{(b)}_{B}$ are, respectively, the
Hamiltonians of the baths for TLS$1$ and TLS$2$. The creation and
annihilation operators $a^{\dag}_{j}$ ($b^{\dag}_{k}$) and $a_{j}$
($b_{k}$) describe the $j$th ($k$th) harmonic oscillator with
frequency $\omega_{aj}$ ($\omega_{bk}$). The interaction Hamiltonian
of the two TLSs with the two IHBs reads
\begin{equation}
H^{(\text{IHB})}_{I}=\sum_{j}g_{1j}(a_{j}^{\dagger}\sigma_{1}^{-}+\sigma
_{1}^{+}a_{j})+\sum_{k}g_{2k}(b_{k}^{\dagger}\sigma
_{2}^{-}+\sigma_{2}^{+}b_{k}).  \label{nondiacouplingindepH}
\end{equation}

On the other hand, in the CHB case, as shown in
Fig.~\ref{schematic}(b), the two TLSs are immersed in a CHB with the
Hamiltonian
\begin{equation}
H^{(\text{CHB}
)}_{B}=\sum_{j}\omega _{aj}a_{j}^{\dagger }a_{j},
\end{equation}
where $a^{\dag}_{j}$ and $%
a_{j}$ are, respectively, the creation and annihilation operators of
the $j$th harmonic oscillator with frequency $\omega_{aj}$. The
interaction Hamiltonian between the two TLSs and the CHB reads as
\begin{equation}
H^{(\text{CHB})}_{I}=\sum_{j}g_{1j}(a_{j}^{\dagger}\sigma_{1}^{-}+\sigma
_{1}^{+}a_{j})+\sum_{j}g_{2j}(a_{j}^{\dagger}\sigma
_{2}^{-}+\sigma_{2}^{+}a_{j}).  \label{nondiacouplingH}
\end{equation}
For simplicity, we have assumed that the TLS-bath coupling strengths
$g_{1j}$, $g_{2j}$, and $g_{2k}$ are real numbers.

\section{\label{Sec:3}Quantum thermalization of two coupled TLSs immersed in
two IHBs}

In this section, we study the quantum thermalization of the two
coupled TLSs that are immersed in two IHBs. We depict the evolution
of the two TLSs in terms of a quantum master equation. By solving
the equations of motion of the density matrix elements to obtain
steady-state solution, we study the steady-state properties of the
two coupled TLSs.

\subsection{Equations of motion and steady-state solutions}

We consider the situation wherein the environments of the two TLSs
are memory-less and the couplings between the TLSs and the
environments are weak. Then, we may adopt the usual Born-Markov
approximation in derivation of quantum master equation. At the same
time, we derive the master equation in eigenstate representation of
the two coupled TLSs so that we can safely make the secular
approximation (equivalent to the rotating-wave approximation) to
obtain a time-independent quantum master equation by neglecting the
rapidly oscillating terms~\cite{Breuer}. Therefore, our discussions
are under the Born-Markov framework.

In the case of two IHBs, the evolution of the two coupled TLSs is governed
by the following Born-Markov master equation in the Schr\"{o}dinger picture,
\begin{eqnarray}
\dot{\rho}_{S}&=&i[\rho_{S},H_{\text{TLSs}}]  \notag \\
&&+\sum_{(i,j)}\left[\Gamma_{ji}(2\tau_{ij}\rho
_{S}\tau_{ji}-\tau_{jj}\rho_{S}-\rho_{S}\tau_{jj})\right.  \notag \\
&&\left.+\Gamma_{ij}(2\tau_{ji}\rho
_{S}\tau_{ij}-\tau_{ii}\rho_{S}-\rho_{S}\tau_{ii})\right]  \notag \\
&&+2\Lambda _{1}(\tau_{42}\rho _{S}\tau_{13} +\tau_{31}\rho _{S}\tau_{24})
\notag \\
&&+2\Lambda _{2}(\tau_{21}\rho _{S}\tau_{34} +\tau_{43}\rho_{S}\tau_{12})
\notag \\
&&+2\Lambda _{3}(\tau_{24}\rho _{S}\tau_{31} +\tau_{13}\rho _{S}\tau_{42})
\notag \\
&&+2\Lambda _{4}(\tau_{12}\rho _{S}\tau_{43} +\tau_{34}\rho _{S}\tau_{21}),
\label{masterequation}
\end{eqnarray}
which will be derived in detail in
Appendix~\ref{appeindependentbath}. In Eq.~(\ref{masterequation}),
$\rho_{S}$ is the reduced density operator of the two TLSs. The
operators $\tau_{ij}$ are defined by
$\tau_{ij}=|\lambda_{i}\rangle\langle\lambda_{j}|$ in the eigenstate
representation of the two coupled TLSs. The summation parameters
$(i,j)$ in the second line of Eq.~(\ref{masterequation}) can take
$(i,j)=(4,2),(3,1),(2,1),$ and $(4,3)$. In the present model, the
effective rates in Eq.~(\ref{masterequation}) are defined as
$\Gamma_{13}=\Gamma_{24}=\Gamma_{1}$, $\Gamma_{31}=\Gamma_{42}=
\Gamma_{2}$, $\Gamma_{12}=\Gamma_{34}=\Gamma_{3}$, and $\Gamma_{21}=
\Gamma_{43}=\Gamma_{4}$, with
\begin{eqnarray}
\Gamma_{1}&=&\cos^{2}(\theta/2)A_{1}(\varepsilon_{1})+\sin^{2}(\theta/2)
B_{1}(\varepsilon_{1}),  \notag \\
\Gamma _{2} &=&\cos^{2}(\theta/2)A_{2}(\varepsilon_{1})+\sin
^{2}(\theta/2)B_{2}(\varepsilon_{1}),  \notag \\
\Gamma _{3} &=&\sin^{2}(\theta/2)A_{1}(\varepsilon_{2})+\cos
^{2}(\theta/2)B_{1}(\varepsilon_{2}),  \notag \\
\Gamma _{4} &=&\sin^{2}(\theta/2) A_{2}(\varepsilon_{2})+\cos
^{2}(\theta/2)B_{2}(\varepsilon_{2}),  \notag \\
\Lambda _{1} &=&\cos^{2}(\theta/2)A_{1}(\varepsilon_{1})-\sin
^{2}(\theta/2)B_{1}(\varepsilon_{1}),  \notag \\
\Lambda _{2} &=&-\sin ^{2}(\theta/2)A_{1}(\varepsilon_{2})+\cos
^{2}(\theta/2)B_{1}(\varepsilon_{2}),  \notag \\
\Lambda _{3} &=&\cos^{2}(\theta/2)A_{2}(\varepsilon_{1}) -\sin
^{2}(\theta/2)B_{2}(\varepsilon_{1}),  \notag \\
\Lambda _{4} &=&-\sin^{2}(\theta/2)A_{2}(\varepsilon_{2})+\cos
^{2}(\theta/2)B_{2}(\varepsilon_{2}),  \label{defofGmmaandLamb}
\end{eqnarray}
where we define $A_{1}(\varepsilon_{i})=\gamma_{a}(\varepsilon
_{i})[\bar{n}_{a}(\varepsilon _{i})+1]$,
$A_{2}(\varepsilon_{i})=\gamma_{a}(\varepsilon
_{i})\bar{n}_{a}(\varepsilon_{i})$,
$B_{1}(\varepsilon_{i})=\gamma_{b}( \varepsilon
_{i})[\bar{n}_{b}(\varepsilon_{i})+1]$, and $B_{2}(
\varepsilon_{i})=\gamma_{b}(\varepsilon
_{i})\bar{n}_{b}(\varepsilon_{i})$ $ (i=1,2)$. The energy
separations $\varepsilon_{1}$ and $\varepsilon_{2}$ are introduced
as $\varepsilon_{1}=E_{\lambda_{1}}-E_{\lambda_{3}}=E_{
\lambda_{2}}-E_{\lambda_{4}}
=\omega_{m}+\sqrt{\Delta\omega^{2}/4+\xi ^{2}}$ and
$\varepsilon_{2}=E_{\lambda_{1}}-E_{\lambda_{2}}=E_{\lambda_{3}}-E_{
\lambda_{4}} =\omega_{m}-\sqrt{\Delta\omega^{2}/4+\xi ^{2}}$. The
rates $ \gamma_{a}(\varepsilon _{i})=\pi\varrho_{a}(\varepsilon
_{i})g^{2}_{1}(\varepsilon _{i})$ and
$\gamma_{b}(\varepsilon_{i})=\pi \varrho_{b}(\varepsilon
_{i})g^{2}_{2}(\varepsilon _{i})$, where $ \varrho_{a}(\varepsilon
_{i})$ and $\varrho_{b}(\varepsilon _{i})$ are, respectively, the
densities of state at energy $\varepsilon _{i}$ of the heat baths
for TLS$1$ and TLS$2$. In the following we assume $
\gamma_{a}(\varepsilon _{1})=\gamma_{b}(\varepsilon
_{1})=\gamma_{1}$ and $ \gamma_{a}(\varepsilon
_{2})=\gamma_{b}(\varepsilon _{2})=\gamma_{2}$. In addition, we
introduce the average thermal excitation numbers $\bar{n}
_{a}(\varepsilon _{i})=1/[\exp(\varepsilon _{i}/T_{1})-1]$ (with the
Boltzmann constant $k_{B}=1$) and $\bar{n} _{b}(\varepsilon
_{i})=1/[\exp(\varepsilon _{i}/T_{2})-1]$ $(i=1,2)$ for the heat
baths of TLS$1$ and TLS$2$, at the respective temperatures $T_{1}$
and $T_{2}$~\cite{Breuer}.

Based on quantum master equation~(\ref{masterequation}), it is
straightforward to obtain optical Bloch equations for the density
matrix elements of the two TLSs in the eigenstate representation.
Denoting
$\mathbf{X}(t)=\left(\langle\tau_{11}(t)\rangle,\langle\tau_{22}(t)\rangle,
\langle\tau_{33}(t)\rangle,\langle\tau_{44}(t)\rangle\right)^{T}$
(``$T$" stands for matrix transpose), then the optical Bloch
equations for the diagonal density matrix elements in the eigenstate
representation can be expressed as
\begin{equation}
\dot{\mathbf{X}}(t)=\mathbf{M}^{\text{(IHB)}}\mathbf{X}(t),
\label{OBEforihb}
\end{equation}
where the coefficient matrix $\mathbf{M}^{\text{(IHB)}}$ is defined by
\begin{equation}
\mathbf{M}^{\text{(IHB)}}=-2\left(
\begin{array}{cccc}
\Gamma _{1}+\Gamma _{3} & -\Gamma _{4} & -\Gamma _{2} & 0 \\
-\Gamma _{3} & \Gamma _{1}+\Gamma _{4} & 0 & -\Gamma _{2} \\
-\Gamma _{1} & 0 & \Gamma _{2}+\Gamma _{3} & -\Gamma _{4} \\
0 & -\Gamma _{1} & -\Gamma _{3} & \Gamma _{2}+\Gamma _{4}
\end{array}
\right).
\end{equation}
From Eq.~(\ref{OBEforihb}), we can see that the evolution of the diagonal
density matrix elements decouples with off-diagonal elements.

The transient solutions of optical Bloch equation~(\ref{OBEforihb})
can be obtained with the Laplace transform method. To study quantum
thermalization, however, it is sufficient to obtain the steady-state
solutions,
\begin{eqnarray}  \label{steadystatesolution}
\langle\tau_{11}\rangle_{ss}&=&\frac{\Gamma_{2}\Gamma_{4}}{
(\Gamma_{1}+\Gamma_{2}) (\Gamma_{3}+\Gamma_{4})},  \notag \\
\langle\tau_{22}\rangle_{ss}&=&\frac{\Gamma_{2}\Gamma_{3}}{
(\Gamma_{1}+\Gamma_{2}) (\Gamma_{3}+\Gamma_{4})},  \notag \\
\langle\tau_{33}\rangle_{ss}&=&\frac{\Gamma_{1}\Gamma _{4}}{(
\Gamma_{1}+\Gamma_{2})(\Gamma _{3}+\Gamma_{4})},  \notag \\
\langle\tau_{44}\rangle_{ss}&=&\frac{\Gamma_{1}\Gamma _{3}}{
(\Gamma_{1}+\Gamma _{2})(\Gamma_{3}+\Gamma_{4})},
\end{eqnarray}
where the subscript ``ss" means steady-state solutions.

According to Eq.~(\ref{masterequation}), we can also obtain the equations of
motion for these off-diagonal density matrix elements of the two TLSs as
follows:
\begin{eqnarray}
\langle \dot{\tau}_{21}(t)\rangle  &=&-(2\Gamma _{1}+\Gamma _{3}+\Gamma
_{4}-i\varepsilon _{2})\langle \tau _{21}(t)\rangle +2\Lambda _{3}\langle
\tau _{43}(t)\rangle ,  \notag \\
\langle \dot{\tau}_{31}(t)\rangle  &=&-(\Gamma _{1}+\Gamma _{2}+2\Gamma
_{3}-i\varepsilon _{1})\langle \tau _{31}(t)\rangle +2\Lambda _{4}\langle
\tau _{42}(t)\rangle ,  \notag \\
\langle \dot{\tau}_{41}(t)\rangle  &=&-(\Gamma _{1}+\Gamma _{2}+\Gamma
_{3}+\Gamma _{4}-i\varepsilon _{1}-i\varepsilon _{2})\langle \tau
_{41}(t)\rangle ,  \notag \\
\langle \dot{\tau}_{32}(t)\rangle  &=&-(\Gamma _{1}+\Gamma _{2}+\Gamma
_{3}+\Gamma _{4}-i\varepsilon _{1}+i\varepsilon _{2})\langle \tau
_{32}(t)\rangle ,  \notag \\
\langle \dot{\tau}_{42}(t)\rangle  &=&-(\Gamma _{1}+\Gamma _{2}+2\Gamma
_{4}-i\varepsilon _{1})\langle \tau _{42}(t)\rangle +2\Lambda _{2}\langle
\tau _{31}(t)\rangle ,  \notag \\
\langle \dot{\tau}_{43}(t)\rangle  &=&-(2\Gamma _{2}+\Gamma
_{3}+\Gamma _{4}-i\varepsilon _{2})\langle \tau _{43}(t)\rangle
+2\Lambda _{1}\langle \tau _{21}(t)\rangle.\nonumber\\
\end{eqnarray}
Other off-diagonal elements can be obtained via $\langle \tau
_{ij}(t)\rangle ^{\ast }=\langle \tau _{ji}(t)\rangle $. It can be
found that the steady-state solutions of the equations of motion for
these off-diagonal elements are zero,
\begin{eqnarray}
\langle \tau _{ij}\rangle _{ss}=0,\hspace{0.5 cm}i\neq
j.\label{sszero}
\end{eqnarray}
Based on the steady-state solutions for these density matrix
elements, we can analyze the steady-state properties of the two
coupled TLSs.

\subsection{Quantum thermalization in eigenstate representation}

For the present system with two IHBs, when the coupling between the
two TLSs is stronger than the TLS-bath couplings, the two coupled
TLSs can be regarded as an effective four-level system connected
with two IHBs. Therefore, in the eigenstate representation, the
dynamic evolution process of the two-coupled TLSs approaching their
steady state of thermal equilibrium can be understood as a
non-equilibrium quantum thermalization: thermalization of a quantum
system connected with many IHBs at different
temperatures~\cite{Jou2003,Zurcher1990,Komatsu2008,Trimper2006,Eckmann1999,Huang2009,Johal2009,Beer}.

As the steady state of the two coupled TLSs is completely mixed in
eigenstate representation, we can introduce effective temperatures to
characterize the relation between any two eigenstates based on their
steady-state populations. From Eq.~(\ref{steadystatesolution}) we can find
the following relations
\begin{eqnarray}
\label{ratios}
\frac{\langle\tau_{11}\rangle_{ss}}{\langle\tau_{22}\rangle_{ss}}&=&\frac{
\langle\tau_{33}\rangle_{ss}}{\langle\tau_{44}\rangle_{ss}}=\frac{\Gamma_{4}
}{\Gamma_{3}},\nonumber\\
\frac{\langle\tau_{11}\rangle_{ss}}{
\langle\tau_{33}\rangle_{ss}}&=&\frac{\langle\tau_{22}\rangle_{ss}}{
\langle\tau_{44}\rangle_{ss}}=\frac{\Gamma_{2}}{\Gamma_{1}}.
\end{eqnarray}
Generally, it is impossible to define an effective temperature for
the effective four-level system at steady state. According to
Eq.~(\ref{ratios}), we can characterize the state of the effective
four-level system via introducing the following two effective
temperatures~\cite{Quan2005} by
\begin{eqnarray}
\frac{\langle\tau_{11}\rangle_{ss}}{\langle\tau_{22}\rangle_{ss}}
&=&\frac{\langle\tau_{33}\rangle_{ss}}{\langle\tau_{44}\rangle_{ss}}
=\exp\left[-\frac{\varepsilon_{2}}{T_{\textrm{eff}}(\varepsilon_{2})}\right],  \notag \\
\frac{\langle\tau_{22}\rangle_{ss}}{\langle\tau_{33}\rangle_{ss}}&
=&\exp\left[-\frac{\varepsilon_{1}-\varepsilon_{2}}{T_{\textrm{eff}}
(\varepsilon_{1}-\varepsilon_{2})}\right].
\end{eqnarray}
In terms of Eq.~(\ref{ratios}), we get
\begin{eqnarray}
T_{\textrm{eff}}(\varepsilon_{1})&=\frac{\varepsilon_{1}}{\ln\left(\Gamma_{1}/
\Gamma_{2}\right)},\hspace{0.5cm}
T_{\textrm{eff}}(\varepsilon_{2})=\frac{
\varepsilon_{2}}{\ln\left(\Gamma_{3}/\Gamma_{4}\right)}.
\end{eqnarray}
When the two IHBs have the same temperatures, namely $T_{1}=T_{2}=T$, we can
show that the above two effective temperatures are equal to those of the two
IHBs, that is
\begin{eqnarray}
T_{\textrm{eff}}(\varepsilon_{1})=T_{\textrm{eff}}(\varepsilon_{2})=T.
\label{equaltemps}
\end{eqnarray}
For the nonequilibrium case of $T_{1}\neq T_{2}$, the two effective
temperatures are different for general case. We find the following
relations $\min(T_{1},T_{2})\leq
T_{\textrm{eff}}(\varepsilon_{1})\leq \max(T_{1},T_{2})$ and
$\min(T_{1},T_{2})\leq T_{\textrm{eff}}(\varepsilon_{2})\leq
\max(T_{1},T_{2})$, which mean that the two effective temperatures
$T_{\textrm{eff}}(\varepsilon_{1})$ and
$T_{\textrm{eff}}(\varepsilon_{2})$ will be within the region from
$\min(T_{1},T_{2})$ to $\max(T_{1},T_{2})$~\cite{Liao}. Under some
special conditions, the two effective temperatures could be equal.
In this case, we consider that the effective four-level system is
thermalized by the non-equilibrium environments: two IHBs at
different temperatures.
\begin{figure}[tbp]
\includegraphics[bb=34 24 360 260, width=3.3 in]{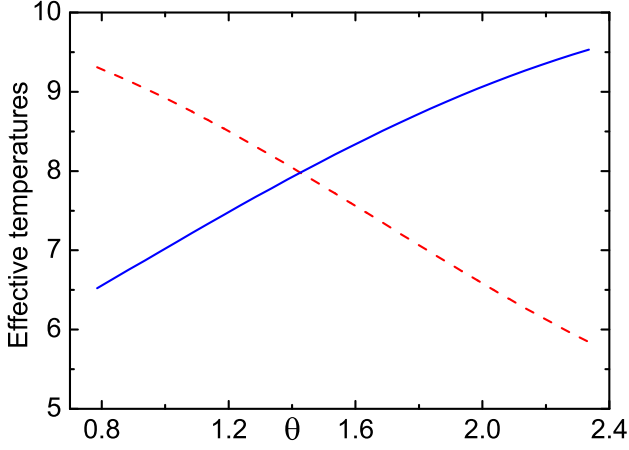}
\caption{(Color online) Plot of the scaled effective temperatures
$T_{\textrm{eff}}(\protect\varepsilon_{1})/\gamma$ (solid blue line)
and $T_{\textrm{eff}}(\protect \varepsilon_{2})/\gamma$ (dashed red
line) vs the mixing angle $\theta$. Other parameters are set as
$\gamma_{1}=\gamma_{2}=\gamma$, $\xi/\gamma=10$,
$\omega_{m}/\gamma=20$, $T_{1}/\gamma=5$, and $T_{2}/\gamma=10$.}
\label{efftemforihb}
\end{figure}

In Fig.~\ref{efftemforihb}, we plot the two effective temperatures
$T_{\textrm{eff}}(\varepsilon_{1})$ and
$T_{\textrm{eff}}(\varepsilon_{2})$ as a function of the mixing
angle $\theta$. It shows that $T_{\textrm{eff}}(\varepsilon_{1})$
and $T_{\textrm{eff}}(\varepsilon_{2})$ can be equal for a special
mixing angle $\theta_{0}$, which is determined by
$\frac{\varepsilon_{1}}{\ln\left(\Gamma_{1}/\Gamma_{2}\right)}
=\frac{\varepsilon_{2}}{\ln\left(\Gamma_{3}/\Gamma_{4}\right)}$.
That is to say the effective four-level system formed by the two
coupled TLSs can be thermalized with an effective temperature when
$\Delta\omega=\xi/\tan\theta_{0}$. Note that this effective
temperature actually is a nonequilibrium
temperature~\cite{Hatano2003}. We can also see from
Fig.~\ref{efftemforihb} that the two effective temperatures
$T_{\textrm{eff}}(\varepsilon_{1})$ and
$T_{\textrm{eff}}(\varepsilon_{2})$ are within the region from
$T_{1}$ to $T_{2}$. We point out that the mixing angle $\theta$
should be chosen to make sure that
$E_{\lambda_{1}}>E_{\lambda_{2}}$.

\subsection{Quantum thermalization in bare-state representation}

When the coupling between the two TLSs is weaker than the TLS-bath
couplings, we understand the quantum thermalization in bare-state
representation. We can express the bare states with the eigenstates
as
\begin{eqnarray}
|\eta_{1}\rangle&=&\vert\lambda_{1}\rangle, \hspace{0.5
cm}|\eta_{2}\rangle
=\cos(\theta/2)\vert\lambda_{2}\rangle-\sin(\theta /2)\vert
\lambda_{3}\rangle,\nonumber\\
|\eta_{3}\rangle&=&\sin(\theta/2)\vert\lambda_{2} \rangle+\cos(
\theta /2)\vert\lambda_{3}\rangle,\hspace{0.5 cm}|\eta_{4}\rangle=
\vert\lambda_{4}\rangle.
\end{eqnarray}
Then we can obtain the relations
\begin{eqnarray}
\label{transforma}
\langle\sigma_{l=1,2}^{z}(t)\rangle&=&\langle\tau
_{11}(t)\rangle-\langle\tau _{44}(t)\rangle  \notag \\
&&-(-1)^{l}\cos\theta[\langle\tau_{22}(t)\rangle-\langle\tau_{33}(t)\rangle]
\notag \\
&&+(-1)^{l}\sin\theta[\langle\tau _{23}(t)\rangle+\langle\tau
_{32}(t)\rangle ].
\end{eqnarray}
According to Eqs.~(\ref{steadystatesolution}),~(\ref{sszero}),
and~(\ref{transforma}), the steady-state solution can be obtained as
\begin{eqnarray}  \label{sigma12ss}
\langle\sigma_{l=1,2}^{z}\rangle_{ss}&=&\frac{(\Gamma_{2}\Gamma
_{4}-\Gamma_{1}\Gamma _{3})-(-1)^{l}\cos\theta(\Gamma_{2}\Gamma
_{3}-\Gamma_{1}\Gamma _{4})}{(\Gamma_{1}+\Gamma_{2})(\Gamma
_{3}+\Gamma_{4})}
.  \notag \\
\end{eqnarray}
In addition, the off-diagonal elements of the density matrices of the two
TLSs can be expressed as
\begin{eqnarray}
\langle \sigma^{+}_{1}(t)\rangle&=&\sin(\theta/2)[\langle
\tau_{14}(t)\rangle-\langle \tau_{34}(t)\rangle]  \notag \\
&&+\cos(\theta/2)[\langle \tau_{13}(t)\rangle+\langle \tau_{24}(t)\rangle],
\notag \\
\langle \sigma^{+}_{2}(t)\rangle&=&\sin(\theta/2)[\langle
\tau_{24}(t)\rangle-\langle \tau_{13}(t)\rangle]  \notag \\
&&+\cos(\theta/2)[\langle \tau_{12}(t)\rangle+\langle \tau_{34}(t)\rangle].
\end{eqnarray}
Because of $\langle \tau_{ij}(t)\rangle_{ss}=0$ $(i\neq j)$, we have
$ \langle \sigma^{+}_{1}\rangle_{ss}=\langle
\sigma^{+}_{2}\rangle_{ss}=0$, which implies that the steady states
of the two TLSs in bare-state representation are completely mixed.
Based on these, it is possible to introduce two effective
temperatures of the two TLSs as follows:
\begin{eqnarray}
\label{Teff}
T_{\textrm{eff}}(\omega_{l})&=&\frac{\omega_{l}}{\ln\left(\frac{
1-\langle\sigma^{z}_{l}\rangle_{ss}}
{1+\langle\sigma^{z}_{l}\rangle_{ss}} \right)},\hspace{0.5 cm}l=1,2.
\end{eqnarray}
\begin{figure}[tbp]
\center
\includegraphics[bb=21 21 368 267, width=3.3 in]{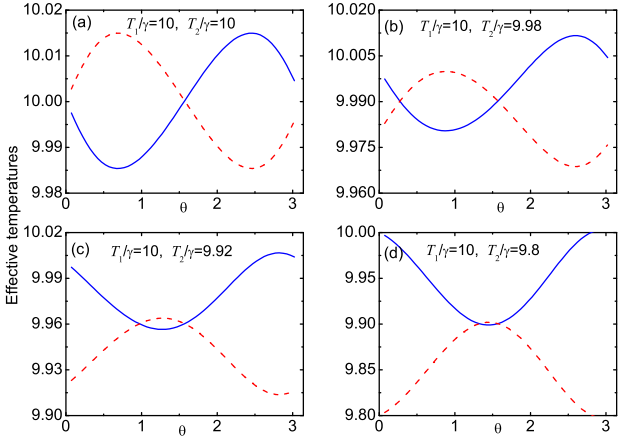}
\caption{(Color online) Plot of the scaled effective temperatures
$T_{\textrm{eff}}(\omega_{1})/\gamma$ (solid blue line) and
$T_{\textrm{eff}}(\omega_{2})/\gamma$ (dashed red line) vs the
mixing angle $\protect\theta$ for various temperature distributions
(shown in figures). Other parameters are set as
$\gamma_{1}=\gamma_{2}=\gamma$, $\xi/\gamma=0.1$, and
$\omega_{m}/\gamma=20$.} \label{efftemforihbbarerep}
\end{figure}

In Fig.~\ref{efftemforihbbarerep}, we plot the effective
temperatures $T_{\textrm{eff}}(\omega_{1})$ and
$T_{\textrm{eff}}(\omega_{2})$ as a function of the mixing angle
$\theta$ for various temperature distributions. From
Fig.~\ref{efftemforihbbarerep}, we can see the following three
interesting results. Firstly, when $\theta\approx\pi/2$,
$T_{\textrm{eff}}(\omega_{1})$ and $ T_{\textrm{eff}}(\omega_{2})$
become approximately equal. Specially, at the resonant point
$\theta=\pi/2$, the two TLSs have the same temperatures no matter
whether the two bath temperatures are the same or not (see the cross
point at $\theta=\pi/2$ in figures). This result can be explained as
follows: When the two TLSs are nearly in resonance with each other,
the dipole-dipole interaction can induce population exchange between
the two TLSs such that their temperatures are approximately equal.
When $\theta=\pi/2$ ($\omega_{1}=\omega_{2}$), Eq.~(\ref{sigma12ss})
reduces to
\begin{eqnarray}
\langle\sigma_{1}^{z} \rangle_{ss}=\langle\sigma
_{2}^{z}\rangle_{ss}=\frac{\Gamma_{2}\Gamma_{4}-
\Gamma_{1}\Gamma_{3}}
{(\Gamma_{1}+\Gamma_{2})(\Gamma_{3}+\Gamma_{4})},
\end{eqnarray}
then we have
$T_{\textrm{eff}}(\omega_{1})=T_{\textrm{eff}}(\omega_{2})$.

Secondly, when $\theta$ is near to $0$ and $\pi$,
$T_{\textrm{eff}}(\omega_{1})$ and $T_{\textrm{eff}}(\omega_{2})$
approach approximately $T_{1}$ and $T_{2}$, respectively. This
result can be understood as follows: Near to $\theta\approx0$ and
$\pi$, the energy detuning $|\Delta\omega|=2\xi/|\tan\theta|$
between the two TLSs is very large, and the population exchange
between them can be neglected for a weak coupling strength $\xi$. At
the same time, the energy separation shifts of the two TLSs are
neglectable [this can be seen from the following effective
Hamiltonian~(\ref{effectiveH})]. Hence, the temperatures of the two
TLSs should be equal to those of their IHBs, respectively.

Thirdly, in Fig.~\ref{efftemforihbbarerep}, there are some regions
of $\theta$ where the effective temperature
$T_{\textrm{eff}}(\omega_{1})$ of TLS$1$ could be smaller than the
temperature $T_{\textrm{eff}}(\omega_{2})$ of TLS$2$ although the
bath temperatures $T_{1}>T_{2}$. This is a counterintuitive result.
Intuitively, in these cases $T_{\textrm{eff}}(\omega_{2})$ should be
smaller than $T_{\textrm{eff}}(\omega_{1})$, because TLS$2$ is
connected with a low-temperature bath while TLS$1$ is connected with
a high-temperature bath [In Fig.~\ref{efftemforihbbarerep}(a), when
the two bath temperatures are the same, the effective temperatures
of the two TLSs should be equal]. Actually, the counterintuitive
result is a net effect of the energy separations of the two TLSs,
the bath temperatures, and the coupling between the two TLSs
(coupling induces population exchange and energy shifts). From
Figs.~\ref{efftemforihbbarerep}(a) to~\ref{efftemforihbbarerep}(d),
the curves change gradually with the increase of the temperature
difference $T_{1}-T_{2}$ between the two baths. And the
counterintuitive region decreases with the increase of the
temperature difference $T_{1}-T_{2}$.

In the following, we present a microscopic explanation for
Fig.~\ref{efftemforihbbarerep} as a physical insight of the
counterintuitive phenomenon. When the values of $\theta$ are far
from the resonant point (not near to $0$ and $\pi$), the two TLSs
will have large detuning from each other. Under the large detuning
condition $\xi\ll\Delta\omega$, the real population exchange between
the two TLSs is compressed, the dipole-dipole interaction induces
energy shift to the two TLSs. In this case, we can derive an
effective Hamiltonian to describe the two TLSs with the
Fr\"{o}hlich-Nakajima transformation
approach~\cite{Frohlich,Nakajima}. Starting from the Hamiltonian
$H_{\textrm{TLSs}}=H'_{0}+H'_{I}$ with $H'_{0}=(\omega_{1}\sigma
_{1}^{z}+\omega_{2}\sigma_{2}^{z})/2$ and $H'_{I}=\xi(\sigma
_{1}^{+}\sigma_{2}^{-}+\sigma_{1}^{-}\sigma_{2}^{+})$, we introduce
an operator
$S=-\xi(\sigma_{1}^{+}\sigma_{2}^{-}-\sigma_{1}^{-}\sigma_{2}^{+})/\Delta\omega$,
which meets the condition $H'_{I}+[H'_{0},S]=0$. Then the effective
Hamiltonian reads as
\begin{eqnarray}
H_{\textrm{eff}}\equiv H'_{0}+\frac{1}{2}[H'_{I},S]
=\frac{1}{2}\bar{\omega}_{1}\sigma
_{1}^{z}+\frac{1}{2}\bar{\omega}_{2}\sigma
_{2}^{z},\label{effectiveH}
\end{eqnarray}
where the shifted energy separations are defined by
$\bar{\omega}_{1}=\omega_{1}+\xi^{2}/\Delta\omega$ and
$\bar{\omega}_{2}=\omega_{2}-\xi^{2}/\Delta\omega$.

We can see from Eq.~(\ref{effectiveH}) that, under the large
detuning condition, there is no effective coupling between the two
TLSs. The dipole-dipole interaction between the two TLSs shifts
their energy separations slightly. Hence, when the two TLSs (with
the shifted energy separation) are thermalized to thermal
equilibrium with their baths, we have the relation
$\exp(-\bar{\omega}_{l}/T_{l})=p^{(l)}_{e}/p^{(l)}_{g}$ ($l=1,2$)
for the TLS$l$ (with shifted energy separation $\bar{\omega}_{l}$,
excited and ground state populations $p^{(l)}_{e}$ and
$p^{(l)}_{g}$) in thermal equilibrium at temperature $T_{l}$, and
then the effective temperatures defined in Eq.~(\ref{Teff}) should
be
\begin{eqnarray}
T_{\textrm{eff}}(\omega_{l})=\frac{\omega_{l}}{\ln(p^{(l)}_{g}/p^{(l)}_{e})}=\frac{\omega_{l}}{\bar{\omega}_{l}}T_{l}.
\end{eqnarray}
From Eq.~(\ref{effectiveH}), we can see that, for a positive $\xi$,
when $0<\theta<\pi /2$, we have $\Delta\omega>0$, then
$\bar{\omega}_{1}>\omega _{1}$ and $\bar{\omega}_{2}<\omega _{2}$.
Hence the effective temperatures
$T_{\textrm{eff}}(\omega_{1})<T_{1}$ and
$T_{\textrm{eff}}(\omega_{2})>T_{2}$. On the other hand, when $\pi
/2<\theta <\pi$, we have $\Delta \omega<0$, then
$\bar{\omega}_{1}<\omega_{1}$ and $\bar{\omega}_{2}>\omega _{2}$.
Hence, the effective temperatures
$T_{\textrm{eff}}(\omega_{1})>T_{1}$ and
$T_{\textrm{eff}}(\omega_{2})<T_{2}$.

According to the above analysis, we can see that, when $T_{1}=T_{2}$
[Fig.~\ref{efftemforihbbarerep}(a)], there will exist a
counterintuitive region. At the same time, when $\theta$ is near to
$0$ and $\pi$, the shifted energy separation $\xi^{2}/\Delta\omega$
approaches zero, then $T_{\textrm{eff}}(\omega_{l})\approx T_{l}$.
Hence, with the increase of the bath temperature difference
$T_{1}-T_{2}$, the difference between the two effective temperatures
also increase, which leads to the counterintuitive region decreases.
These results can be seen from Fig.~\ref{efftemforihbbarerep}.

In fact, the above intuitive result is based on the phenomenological
master equation
\begin{eqnarray}
\dot{\rho}_{S}&=&i[\rho_{S},H_{\text{TLSs}}]
+\mathcal{L}_{1}[\rho_{S}]+\mathcal{L}_{2}[\rho_{S}],
\label{phenomenologicalmeq}
\end{eqnarray}
with
\begin{eqnarray}
\mathcal{L}_{l=1,2}[\rho_{S}]&=&\frac{\gamma_{l}}{2}(\bar{n} _{l}+1)(
2\sigma_{l}^{-}\rho\sigma_{l}^{+}-\sigma_{l}^{+}\sigma _{l}^{-}\rho -\rho
\sigma_{l}^{+}\sigma_{l}^{-})  \notag \\
&&+\frac{\gamma_{l}}{2}\bar{n}_{l}(2\sigma_{l}^{+}\rho\sigma
_{l}^{-}-\sigma_{l}^{-}\sigma_{l}^{+}\rho-\rho\sigma
_{l}^{-}\sigma_{l}^{+}).
\end{eqnarray}
The superoperator $\mathcal{L}_{l}[\rho_{S}]$ describes the
dissipation of a TLS$l$ ($l=1,2$) immersed in a heat bath at
temperature $T_{l}$ ($\bar{n}_{l}=1/[\exp(\omega_{l}/T_{l})-1]$).
Therefore, Eq.~(\ref{phenomenologicalmeq}) is not valid in the case
of two coupled TLSs, especially when the coupling between the two
TLSs is stronger than the TLS-bath couplings. In addition, in
Eq.~(\ref{phenomenologicalmeq}), the effects on the TLSs from the
two baths are different, one is direct and the other is indirect.
For example, the bath of TLS$1$ affects TLS$1$ directly, while the
bath of TLS$2$ affects TLS$1 $ indirectly through TLS$2$. On the
contrary, our results given in Eq.~(\ref{Teff}) are based on quantum
master equation~(\ref{masterequation}), which is rigorously derived
in the eigen-representation of the two coupled TLSs. Hence, the
dissipation is depicted in the eigen-representation of the two
coupled TLSs. In other words, the two TLSs play equivalent roles and
the two baths directly affect the TLSs. The resonant case is a clear
example for the equivalent role of the two TLSs. In the resonant
case $\theta=\pi/2$, we obtain
$T_{\textrm{eff}}(\omega_{1})=T_{\textrm{eff}}(\omega_{2})$ in terms
of quantum master equation~(\ref{masterequation}), while we get
$T_{\textrm{eff}}(\omega_{1})>T_{\textrm{eff}}(\omega_{2})$ from
Eq.~(\ref{phenomenologicalmeq}) when $T_{1}>T_{2}$.

\subsection{Steady-state entanglement between the two TLSs}

In the IHB case, there exists a dipole-dipole interaction between
the two TLSs. Therefore, a natural question is: what the quantum
entanglement is between the two TLSs after they are thermalized. As
we know, during the thermalization processes (not
antithermalization), all of the initial information of the two
coupled TLSs is totally erased, and the steady state of the two TLSs
is determined by the decay rates and bath temperatures. Hence, we
need to know the steady-state entanglement in the two TLSs. We note
that entanglement dynamics in similar systems has been
studied~\cite{Subrahmanyam}. In the following we apply the
concurrence to quantify the steady-state entanglement in the two
TLSs.

For a $2\times 2$ quantum system (two TLSs) with a density matrix
$\rho $ expressed in the bare-state representation, its
concurrence~\cite{Wootters} is defined as
\begin{equation}
C(\rho )=\max
\{0,\sqrt{s_{1}}-\sqrt{s_{2}}-\sqrt{s_{3}}-\sqrt{s_{4}}\},
\end{equation}
where $s_{i}$ ($i=1,2,3,4$) are the eigenvalues ($s_{1}$ being the
largest one) of the matrix $\rho \tilde{\rho}$. The operator
$\tilde{\rho}$ is defined as
\begin{equation}
\tilde{\rho}=(\sigma_{1}^{y}\otimes
\sigma_{2}^{y})\rho^{\ast}(\sigma_{1}^{y}\otimes\sigma_{2}^{y}),
\end{equation}
where $\rho ^{\ast }$ is the complex conjugate of $\rho$ and
$\sigma_{l}^{y}$ is the Pauli matrix of TLS$l$. For the $2\times 2$
system, $C=0$ and $C=1$ mean, respectively, the density matrix
$\rho$ is an unentangled state and a maximally entangled state. In
particular, for the so-called ``X"-class state with the density
matrix (expressed in the bare-state representation)
\begin{eqnarray}
\rho=\left(
       \begin{array}{cccc}
         \rho_{11} &0 & 0 & \rho_{14} \\
         0 & \rho_{22} & \rho_{23} & 0 \\
         0 & \rho_{32} & \rho_{33} & 0 \\
         \rho_{41} & 0 & 0 & \rho_{44} \\
       \end{array}
     \right),
\end{eqnarray}
the concurrence is~\cite{Ikram}
\begin{eqnarray}
C(\rho)=2\max\left\{0,|\rho_{23}|-\sqrt{\rho_{11}\rho_{44}},|\rho_{14}|-\sqrt{\rho_{22}\rho_{33}}\right\}.\label{Xstateconcu}
\end{eqnarray}
\begin{figure}[tbp]
\includegraphics[bb=19 11 359 260, width=3.3 in]{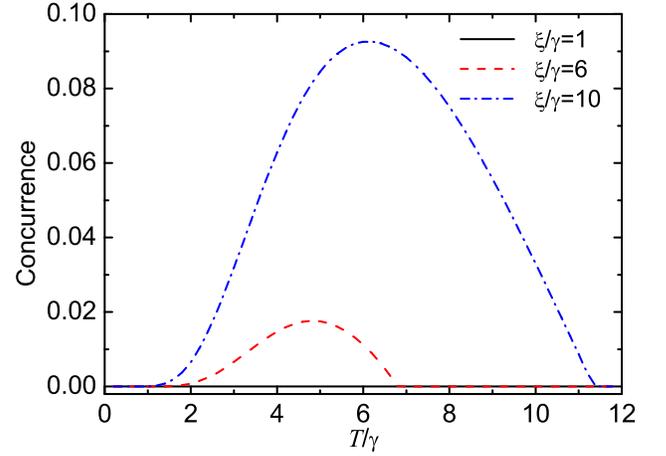}
\caption{(Color online) Plot of the steady-state concurrence
$C(\rho_{ss})$ vs the scaled bath temperature $T/\gamma$ for various
values of $\xi/\gamma$. Other parameters are set as
$\gamma_{1}=\gamma_{2}=\gamma$, $\theta=\pi/2$, $T_{1}=T_{2}=T$, and
$\omega_{m}/\gamma=20$.} \label{ssentangihb}
\end{figure}

Now, for the present system, its density matrix elements in
bare-state representation can be expressed as $\langle\eta _{j}|\rho
|\eta_{i}\rangle=\textrm{Tr}[|\eta_{i}\rangle\langle\eta_{j}|\rho
]=\textrm{Tr}[\mu_{ij}\rho]=\langle\mu_{ij}\rangle$ with the
transition operator $\mu_{ij}=|\eta_{i}\rangle \langle \eta _{j}|$.
The density matrix elements in the eigenstate representation are
expressed by $\langle\lambda_{j}|\rho|\lambda_{i}\rangle
=\textrm{Tr}[|\lambda_{i}\rangle\langle\lambda_{j}|\rho]
=\textrm{Tr}[\tau_{ij}\rho]=\langle\tau_{ij}(t)\rangle$ with $\tau
_{ij}=|\lambda _{i}\rangle \langle \lambda _{j}|$. Since the
concurrence is defined in the bare-state representation, and the
evolution of the system is expressed in the eigenstate
representation. Therefore we need to obtain the transformation
between the two representations as follows:
\begin{eqnarray}
\langle\mu_{11}(t)\rangle&=&\langle\tau_{11}(t)\rangle,\hspace{0.5cm}\langle\mu
_{44}(t)\rangle=\langle\tau_{44}(t)\rangle,  \notag \\
\langle\mu_{22}(t)\rangle&=&\cos^{2}(\theta/2)\langle\tau_{22}(t)\rangle+\sin^{2}(\theta/2)
\langle\tau_{33}(t)\rangle,\nonumber\\
&&-\frac{1}{2}\sin \theta[\langle\tau_{23}(t)\rangle
+\langle\tau_{32}(t)\rangle],\notag \\
\langle\mu_{33}(t)\rangle&=&\sin^{2}(\theta/2)\langle
\tau_{22}(t)\rangle +\cos^{2}(\theta/2)
\langle\tau_{33}(t)\rangle,\notag \\
&&+\frac{1}{2}\sin \theta[\langle\tau_{23}(t)\rangle
+\langle\tau_{32}(t)\rangle],\label{reptransdiag}
\end{eqnarray}
and
\begin{eqnarray}
\langle\mu_{23}(t)\rangle &=&-\sin^{2}(\theta/2)\langle \tau
_{32}(t)\rangle+\cos^{2}(\theta/2)\langle\tau_{23}(t)\rangle, \notag \\
&&+\frac{1}{2}\sin \theta[\langle\tau_{22}(t)\rangle
-\langle\tau_{33}(t)\rangle],\notag \\
\langle\mu_{12}(t)\rangle&=&\cos(\theta/2)\langle \tau
_{12}(t)\rangle-\sin(\theta /2)\langle\tau _{13}(t)\rangle,\notag \\
\langle\mu_{13}(t)\rangle &=&\sin(\theta/2)\langle \tau
_{12}(t)\rangle +\cos (\theta /2)\langle \tau _{13}(t)\rangle,\notag \\
\langle\mu_{14}(t)\rangle&=&\langle\tau_{14}(t)\rangle,\notag\\
\langle\mu_{24}(t)\rangle &=&\cos(\theta/2)\langle \tau
_{24}(t)\rangle-\sin(\theta/2) \langle \tau
_{34}(t)\rangle,\notag \\
\langle\mu_{34}(t)\rangle &=&\sin(\theta/2) \langle \tau
_{24}(t)\rangle+\cos(\theta/2) \langle \tau
_{34}(t)\rangle.\label{reptransoffdiag}
\end{eqnarray}

According to the steady-state solutions given in
Eqs.~(\ref{steadystatesolution}) and~(\ref{sszero}), the
steady-state density matrix of the two TLSs in the bare-state
representation can be obtained with the following nonzero elements
\begin{eqnarray}
\langle\mu_{11}\rangle_{ss}&=&\langle\tau_{11}\rangle_{ss},
\hspace{0.5 cm}
\langle \mu_{44}\rangle_{ss}=\langle\tau_{44}\rangle_{ss},\notag \\
\langle \mu_{22}\rangle_{ss}
&=&\cos^{2}(\theta/2)\langle\tau_{22}\rangle_{ss}+\sin
^{2}(\theta/2)\langle\tau_{33}\rangle_{ss},\notag\\
\langle\mu_{33}\rangle_{ss}&=&\sin^{2}(\theta/2)\langle\tau_{22}\rangle_{ss}+\cos
^{2}(\theta/2)\langle\tau_{33}\rangle_{ss},\notag\\
\langle \mu _{23}\rangle _{ss} &=&\langle \mu _{32}\rangle _{ss}
=\frac{1}{2}\sin\theta(\langle\tau_{22}\rangle_{ss}-\langle\tau_{33}\rangle_{ss}).\label{nonzeelemens}
\end{eqnarray}
The concurrence of this steady state is
\begin{equation}
C(\rho_{ss})=2\max \left\{0,|\langle \mu
_{32}\rangle_{ss}|-\sqrt{\langle \mu_{11}\rangle_{ss}\langle \mu
_{44}\rangle_{ss}}\right\}.\label{ssconcu}
\end{equation}

In the following, we study the steady-state concurrence of the two
coupled TLSs in the case of $\theta=\pi/2$ and $T_{1}=T_{2}=T$. In
Fig.~\ref{ssentangihb}, we plot the steady-state concurrence
$C(\rho_{ss})$ as a function of the bath temperature $T$ for various
values of the dipole-dipole interaction strength $\xi$.
Figure~\ref{ssentangihb} shows that a larger steady-state
concurrence can be created for a larger $\xi$. We also see the
sudden death of the concurrence when the bath temperature increases
up to a critical value. Notice that the phenomenon of threshold
temperature has also been found in thermal entanglement of spin
model~\cite{Quiroga2007,Sinaysky2008,Vedral2001,Wang2002}.

\section{\label{Sec:4}Quantum thermalization of two coupled TLSs immersed in a CHB}

In the above section, we have studied the quantum thermalization of two
coupled TLSs immersed in two IHBs. However, in some cases, the two coupled
TLSs can be considered to be placed in a CHB. In this section, we shall
study the quantum thermalization of two coupled TLSs immersed in a CHB.

\subsection{\label{Subsec:4-1}Equations of motion and steady-state solutions}

The quantum master equation describing the evolution of the two TLSs
immersed in a CHB at temperature $T$ has the same form as
Eq.~(\ref{masterequation}), but the effective rates are not the same
as those in the IHB case. In the CHB case, the rates read as
\begin{eqnarray}
\Gamma_{12}&=&\left[\sin(\theta/2)\sqrt{\gamma_{1}(\varepsilon_{2})}
+\cos(\theta/2)\sqrt{\gamma_{2}(\varepsilon_{2})}\right]^{2}[\bar{n}
(\varepsilon_{2})+1],  \notag \\
\Gamma_{21}&=&\left[\sin(\theta/2)\sqrt{\gamma_{1}(\varepsilon_{2})}
+\cos(\theta/2)\sqrt{\gamma_{2}(\varepsilon_{2})}\right]^{2}\bar{n}
(\varepsilon_{2}),  \notag \\
\Gamma_{13}&=&\left[\cos(\theta/2)\sqrt{\gamma_{1}(\varepsilon_{1})}
-\sin(\theta/2)\sqrt{\gamma_{2}(\varepsilon_{1})}\right]^{2}[\bar{n}
(\varepsilon_{1})+1],  \notag \\
\Gamma_{31}&=&\left[\cos(\theta/2)\sqrt{\gamma_{1}(\varepsilon_{1})}
-\sin(\theta/2)\sqrt{\gamma_{2}(\varepsilon_{1})}\right]^{2}\bar{n}
(\varepsilon_{1}),  \notag \\
\Gamma_{24}&=&\left[\cos(\theta/2)\sqrt{\gamma_{1}(\varepsilon_{1})}
+\sin(\theta/2)\sqrt{\gamma_{2}(\varepsilon_{1})}\right]^{2}[\bar{n}
(\varepsilon_{1})+1],  \notag \\
\Gamma_{42}&=&\left[\cos(\theta/2)\sqrt{\gamma_{1}(\varepsilon_{1})}
+\sin(\theta/2)\sqrt{\gamma_{2}(\varepsilon_{1})}\right]^{2}\bar{n}
(\varepsilon_{1}),  \notag \\
\Gamma_{34}&=&\left[\sin(\theta/2)\sqrt{\gamma_{1}(\varepsilon_{2})}
-\cos(\theta/2)\sqrt{\gamma_{2}(\varepsilon_{2})}\right]^{2}[\bar{n}
(\varepsilon_{2})+1],  \notag \\
\Gamma_{43}&=&\left[\sin(\theta/2)\sqrt{\gamma_{1}(\varepsilon_{2})}
-\cos(\theta/2)\sqrt{\gamma_{2}(\varepsilon_{2})}\right]^{2}\bar{n}
(\varepsilon_{2}),  \notag \\
\Lambda_{1}&=&[\cos^{2}(\theta/2)\gamma_{1}(\varepsilon_{1})-\sin^{2}(\theta/2)\gamma_{2}(
\varepsilon_{1})][\bar{n}(\varepsilon_{1})+1],  \notag \\
\Lambda_{2}&=&[-\sin^{2}(\theta/2)\gamma_{1}(\varepsilon_{2})+\cos^{2}(\theta/2)\gamma_{2}(
\varepsilon_{2})][\bar{n}(\varepsilon_{2})+1],  \notag \\
\Lambda_{3}&=&[\cos ^{2}(\theta/2)\gamma_{1}(\varepsilon_{1})
-\sin^{2}(\theta/2)\gamma_{2}(\varepsilon_{1})]\bar{n}(\varepsilon_{1}),\notag \\
\Lambda_{4}&=&[-\sin^{2}(\theta/2)\gamma_{1}(\varepsilon_{2})
+\cos^{2}(\theta/2)\gamma_{2}(\varepsilon_{2})]\bar{n}(\varepsilon_{2}),
\label{ratesforcommonbath}
\end{eqnarray}
where we introduce the rates $\gamma_{1}(\varepsilon
_{i})=\pi\varrho(\varepsilon _{i})g^{2}_{1}(\varepsilon _{i})$, $%
\gamma_{2}(\varepsilon _{i})=\pi\varrho(\varepsilon
_{i})g^{2}_{2}(\varepsilon _{i})$, and the average thermal excitation number
$\bar{n}(\varepsilon_{i})=1/[\exp(\varepsilon_{i}/T)-1]$ ($i=1,2$). The
detailed derivation of these rates will be given in Appendix~\ref%
{appecommontbath}. In comparison with the rates in the IHB case, the rates
in the CHB case have correlation terms which are induced by the CHB. In the
following discussions, we assume $\gamma_{1}(\varepsilon
_{i})=\gamma_{2}(\varepsilon _{i})\equiv\gamma(\varepsilon _{i})$.

Correspondingly, the optical Bloch equation for the CHB case has the same
form as Eq.~(\ref{OBEforihb}), but the coefficient matrix is replaced by $%
\mathbf{M}^{\text{(CHB)}}$ with the following expression
\begin{equation}
\mathbf{M}^{\text{(CHB)}}=-2\left(
\begin{array}{cccc}
\Gamma_{12}+\Gamma_{13} & -\Gamma_{21} & -\Gamma_{31} & 0 \\
-\Gamma_{12} & \Gamma_{21}+\Gamma_{24} & 0 & -\Gamma_{42} \\
-\Gamma_{13} & 0 & \Gamma_{31}+\Gamma_{34} & -\Gamma_{43} \\
0 & -\Gamma_{24} & -\Gamma_{34} & \Gamma_{42}+\Gamma_{43}%
\end{array}
\right).
\end{equation}
Similar to the IHB case, the equations of motion for diagonal
density matrix elements decouple with the off-diagonal elements. The
steady-state solutions of the present optical Bloch equation read as
\begin{eqnarray}
\label{ssforcommonbath}
\langle \tau _{11}\rangle_{ss} &=&\frac{(\Gamma_{21}+\Gamma_{24}) \Gamma
_{31}\Gamma _{43}+(\Gamma _{31}+\Gamma _{34}) \Gamma _{21}\Gamma _{42}}{A},
\notag \\
\langle \tau_{22}\rangle_{ss} &=&\frac{ (\Gamma _{12}+\Gamma _{13})\Gamma
_{34}\Gamma _{42} +(\Gamma _{42}+\Gamma _{43})\Gamma _{12}\Gamma _{31}}{A},
\notag \\
\langle \tau _{33}\rangle_{ss} &=&\frac{ (\Gamma _{12}+\Gamma _{13})\Gamma
_{43}\Gamma _{24} +(\Gamma _{42}+\Gamma _{43})\Gamma _{21}\Gamma _{13}}{A},
\notag \\
\langle \tau _{44}\rangle_{ss} &=&\frac{(\Gamma _{21}+\Gamma _{24})\Gamma
_{13} \Gamma _{34}+(\Gamma _{31}+\Gamma _{34}) \Gamma _{12}\Gamma _{24}}{A},
\end{eqnarray}
with $A=(\Gamma _{12}+\Gamma
_{13})(\Gamma_{34}\Gamma_{42}+\Gamma_{43}\Gamma_{24}) +(\Gamma _{21}+\Gamma
_{24})(\Gamma _{31}\Gamma_{43}+\Gamma _{13}\Gamma_{34}) +(\Gamma
_{31}+\Gamma _{34})(\Gamma _{21}\Gamma_{42}+\Gamma _{12}\Gamma_{24})
+(\Gamma _{42}+\Gamma _{43})(\Gamma _{21}\Gamma_{13}+\Gamma _{12}\Gamma_{31})
$.

We can also obtain the equations of motion for these off-diagonal density
matrix elements as follows:
\begin{eqnarray}
\langle\dot{\tau}_{21}(t)\rangle&=&-(\Gamma_{21}+\Gamma_{12}+\Gamma
_{13}+\Gamma _{24}-i\varepsilon_{2})\langle \tau _{21}(t)\rangle  \notag \\
&&+2\Lambda_{3}\langle\tau_{43}(t)\rangle,  \notag \\
\langle\dot{\tau}_{31}(t)\rangle&=&-(\Gamma _{12}+\Gamma
_{31}+\Gamma_{13}+\Gamma_{34}-i\varepsilon_{1})\langle \tau _{31}(t)\rangle
\notag \\
&&+2\Lambda _{4}\langle\tau_{42}(t)\rangle,  \notag \\
\langle \dot{\tau}_{41}(t)\rangle&=&-(\Gamma_{12}+\Gamma
_{13}+\Gamma_{42}+\Gamma_{43}-i\varepsilon_{1}-i\varepsilon _{2})
\langle\tau_{41}(t)\rangle,  \notag \\
\langle\dot{\tau}_{32}(t)\rangle&=&-(\Gamma_{21}+\Gamma_{31}+\Gamma
_{24}+\Gamma_{34}-i\varepsilon_{1}+i\varepsilon_{2})\langle \tau
_{32}(t)\rangle,  \notag \\
\langle\dot{\tau}_{42}(t)\rangle&=&-(\Gamma_{21}+\Gamma _{42}+\Gamma
_{24}+\Gamma _{43}-i\varepsilon _{1})\langle \tau _{42}(t)\rangle  \notag \\
&&+2\Lambda _{2}\langle\tau_{31}(t)\rangle,  \notag \\
\langle\dot{\tau}_{43}(t)\rangle&=&-(\Gamma_{31}+\Gamma_{42}+\Gamma
_{43}+\Gamma_{34}-i\varepsilon _{2})\langle\tau _{43}(t)\rangle  \notag \\
&& +2\Lambda_{1}\langle\tau_{21}(t)\rangle.
\end{eqnarray}
The equations of motion for other elements can be obtained by
$\langle\tau_{ij}(t)\rangle=\langle\tau^{\ast}_{ji}(t)\rangle$.
Obviously, the steady-state solutions of these off-diagonal density
matrix elements are zero.

\subsection{Quantum thermalization in eigenstate representation}

Differently from the IHB case, for the present four-level system, we
introduce three effective temperatures to characterize its state.
The three effective temperatures $T_{12}$, $T_{13}$, and $T_{34}$
are defined according to the populations of the four levels as
follows:
\begin{eqnarray}
T_{ij}&=&\frac{E_{\lambda_{i}}-E_{\lambda_{j}}}{\ln\left(\frac{\langle \tau
_{jj}\rangle_{ss}} {\langle \tau _{ii}\rangle_{ss}}\right)}.
\end{eqnarray}
We can show that the above introduced temperatures are the same as that of
the CHB,
\begin{eqnarray}
T_{12}=T_{13}=T_{34}=T,  \label{equaltempeforcomba}
\end{eqnarray}
which means the two coupled TLSs can approach a thermal equilibrium with the
CHB. In other words, the two coupled TLSs in eigenstate representation can
be thermalized by the CHB.

According to Eqs.~(\ref{equaltemps}) and (\ref{equaltempeforcomba}), we know
that when the temperatures of the heat baths are $T$, irrespective of two
IHBs or a CHB, the effective four-level system formed by the two coupled
TLSs can be thermalized into a thermal equilibrium state with the same
temperature $T$. In other words, based on the thermal equilibrium state at
temperature $T$, we can not know whether the two coupled TLSs are connected
with two IHBs or a CHB.

\subsection{Quantum thermalization in bare-state representation}

\begin{figure}[tbp]
\includegraphics[bb=5 23 356 259, width=3.3 in]{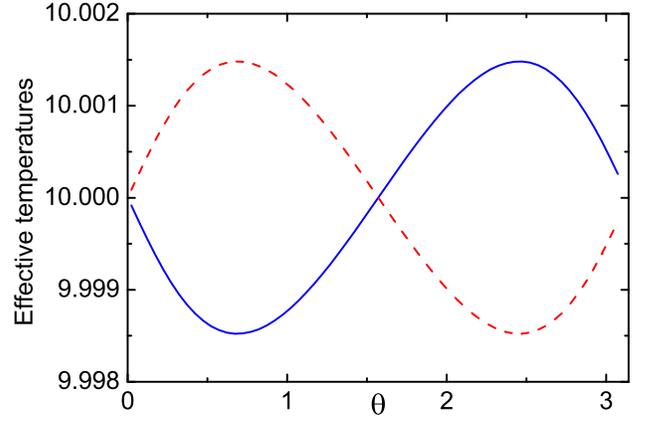}
\caption{(Color online) Plot of the scaled effective temperatures
$T_{\textrm{eff}}(\omega_{1})/\gamma$ (solid blue line) and
$T_{\textrm{eff}}(\omega_{2})/\gamma$ (dashed red line) vs the
mixing angle $\theta$. Other parameters are set as
$\gamma(\varepsilon_{1})=\gamma(\varepsilon_{2})=\gamma$,
$\xi/\gamma=0.1$, $\omega_{m}/\gamma=20$, and $T/\gamma=10$. The
resonant point $\theta=\pi/2$ is useless.} \label{efftemforchb}
\end{figure}

We also investigate the quantum thermalization of the two TLSs in the
bare-state representation. In terms of Eqs.~(\ref{transforma}) and~(\ref%
{ssforcommonbath}), we can obtain the steady-state average values of the two
Pauli operators $\sigma^{z}_{1}$ and $\sigma^{z}_{2}$ as follows:
\begin{eqnarray}
\langle\sigma_{l=1,2}^{z}\rangle_{ss}&=&\frac{(\Gamma_{21}+\Gamma_{24})(%
\Gamma _{31}\Gamma _{43}-\Gamma _{13} \Gamma _{34})}{A}  \notag \\
&&+\frac{(\Gamma _{31}+\Gamma _{34})(\Gamma _{21}\Gamma _{42}-\Gamma
_{12}\Gamma _{24})}{A}  \notag \\
&&+(-1)^{l-1}\cos\theta\left[\frac{(\Gamma _{12}+\Gamma _{13})(\Gamma
_{34}\Gamma _{42}-\Gamma _{43}\Gamma _{24})}{A}\right.  \notag \\
&&\left.+\frac{(\Gamma _{42}+\Gamma _{43})(\Gamma _{12}\Gamma_{31}-\Gamma
_{21}\Gamma_{13})}{A}\right].
\end{eqnarray}
Moreover, we have $\langle \sigma^{+}_{1}\rangle_{ss}=\langle
\sigma^{+}_{2}\rangle_{ss}=0$. Similar to Eq.~(\ref{Teff}) in the above
section, we also introduce two effective temperatures to characterize the
state of the two TLSs. In Fig.~\ref{efftemforchb}, we plot the two effective
temperatures as a function of the mixing angle $\theta$. We emphasize that
the resonant point $\theta=\pi/2$ in Fig.~\ref{efftemforchb} should be taken
out.

We can draw a conclusion from Fig.~\ref{efftemforchb} that the TLS
with a larger energy separation can be thermalized to a thermal
equilibrium state with a lower temperature. This can be seen as
follows: When $ \omega_{1}>\omega_{2}$, we have $0<\theta<\pi/2$,
from Fig.~\ref{efftemforchb} it is clear that the temperature of
TLS$1$ is lower than that of TLS$2$. On the other hand, when
$\omega_{1}<\omega_{2}$, we have $ \pi/2<\theta<\pi$,
Fig.~\ref{efftemforchb} indicates that the temperature of TLS$2$ is
lower than that of TLS$1$ in this region. The physical explanation
for Fig.~\ref{efftemforchb} is the same as that for
Fig.~\ref{efftemforihbbarerep}(a).

\subsection{Quantum anti-thermalization in the resonant case}

When $\theta=\pi/2$, the two TLSs are in resonance, and then the
decay rates $\Gamma_{13}$, $\Gamma_{31}$, $\Gamma_{34}$, and
$\Gamma_{43}$ are zero under the assumption
$\gamma_{1}(\varepsilon_{i}) =\gamma_{2}(\varepsilon_{i})$. Hence,
the eigenstate $|\lambda_{3}\rangle$ decouples with other
eigenstates, resulting in an anti-thermalization phenomenon. The
state $|\lambda_{3}\rangle$ is called as a ``dark
state"~\cite{Gea-Banacloche}. In this case, we need to rewrite new
optical Bloch equations for these density matrix elements. We obtain
the equations of motion for diagonal density matrix elements
\begin{eqnarray}
\langle \dot{\tau}_{11}(t)\rangle &=&-2\Gamma _{12}\langle \tau
_{11}(t)\rangle +2\Gamma _{21}\langle \tau _{22}(t)\rangle,  \notag \\
\langle \dot{\tau}_{22}(t)\rangle &= &2\Gamma _{12}\langle \tau
_{11}(t)\rangle-2(\Gamma_{21}+\Gamma_{24}) \langle \tau_{22}(t)\rangle
\notag \\
&&+2\Gamma _{42}\langle\tau_{44}(t)\rangle  \notag \\
\langle \dot{\tau}_{33}(t)\rangle &=&0,  \notag \\
\langle \dot{\tau}_{44}(t)\rangle &=&2\Gamma _{24}\langle \tau
_{22}(t)\rangle-2\Gamma_{42}\langle\tau_{44}(t)\rangle,
\end{eqnarray}
and for off-diagonal density matrix elements
\begin{eqnarray}
\langle\dot{\tau}_{21}(t)\rangle&=&-(\Gamma_{21}+\Gamma_{12}+\Gamma
_{24}-i\varepsilon _{2})\langle\tau_{21}(t)\rangle,  \notag \\
\langle \dot{\tau}_{31}(t)\rangle &=&-(\Gamma _{12}-i\varepsilon
_{1})\langle \tau _{31}(t)\rangle,  \notag \\
\langle \dot{\tau}_{41}(t)\rangle&=&-(\Gamma_{12}+\Gamma _{42}-i\varepsilon
_{1}-i\varepsilon _{2})\langle\tau _{41}(t)\rangle,  \notag \\
\langle\dot{\tau}_{32}(t)\rangle&=&-(\Gamma_{21}+\Gamma _{24}-i\varepsilon
_{1}+i\varepsilon _{2}) \langle \tau _{32}(t)\rangle,  \notag \\
\langle \dot{\tau}_{42}(t)\rangle&=&-(\Gamma_{21}+\Gamma
_{42}+\Gamma_{24}-i\varepsilon _{1})\langle\tau_{42}(t)\rangle,  \notag \\
\langle \dot{\tau}_{43}(t)\rangle&=&-(\Gamma _{42}-i\varepsilon
_{2})\langle\tau_{43}(t)\rangle.
\end{eqnarray}
The steady-state solutions for these density matrix elements are
\begin{eqnarray}  \label{ssforantitherma}
\langle\tau_{11}\rangle_{ss}&=&\frac{[1-\langle\tau_{33}(0)\rangle]
\Gamma_{21}\Gamma_{42}}{\Gamma_{12}\Gamma_{42}+\Gamma_{12}\Gamma_{42}+
\Gamma_{21}\Gamma_{42}},\notag \\
\langle\tau_{22}\rangle_{ss}&=&\frac{[1-\langle\tau_{33}(0)\rangle]
\Gamma_{12}\Gamma_{42}}{\Gamma_{12}\Gamma_{42}+\Gamma_{12}\Gamma_{42}+
\Gamma_{21}\Gamma_{42}},\notag \\
\langle\tau_{33}\rangle_{ss}&=&\langle\tau_{33}(0)\rangle,\notag \\
\langle\tau_{44}\rangle_{ss}&=&\frac{[1-\langle\tau_{33}(0)\rangle]
\Gamma_{12}\Gamma_{24}}{\Gamma_{12}\Gamma_{42}+\Gamma_{12}\Gamma_{42}+
\Gamma_{21}\Gamma_{42}},\notag \\
\langle\tau_{ij}\rangle_{ss}&=&0,\hspace{0.5 cm}i\neq j,
\end{eqnarray}
where $\langle\tau_{33}(0)\rangle=0$ is the initial population of
state $ |\lambda_{3}\rangle$. It is obvious that the steady state of
the two coupled TLSs depends on its initial state. Therefore, the
two coupled TLSs exhibit a phenomenon of anti-thermalization in the
sense that the heat bath can not erase totally the initial
information of the two TLSs. For example, when initially the two
coupled TLSs are prepared in state $|\lambda_{3}\rangle$, they will
stay in $|\lambda_{3}\rangle$ forever. However, in the subspace
spanned by the three eigenstates $|\lambda_{1}\rangle$,
$|\lambda_{2}\rangle$ , and $|\lambda_{4}\rangle$, the initial
information of the two coupled TLSs can be totally erased, which can
also be seen from Eq.~(\ref{ssforantitherma} ) when
$\langle\tau_{33}(0)\rangle=0$.

\subsection{Steady-state entanglement between the two TLSs}

In the CHB case, in addition to the dipole-dipole interaction
between the two TLSs, the common bath can also provide a physical
mechanism to entangle the two TLSs. According to
Eqs.~(\ref{reptransdiag}) and~(\ref{reptransoffdiag}),  the
steady-state density matrix elements of the two TLSs in the CHB case
have the same form as those given in Eq.~(\ref{nonzeelemens}).
However, now the steady-state solutions of the eigenstate
populations $\langle\tau_{jj}\rangle_{ss}$ are given by
Eq.~(\ref{ssforcommonbath}). Correspondingly, we can obtain the
concurrence between the two TLSs in terms of Eq.~(\ref{ssconcu}).
For a given nonresonant $\theta$, the figure in the CHB case is very
similar to Fig.~\ref{ssentangihb} (so it is not shown here). The
phenomenon of threshold temperature also exists in the CHB case.

\section{\label{Sec:5}conclusion and discussions}

In conclusion, we have studied the quantum thermalization of two
coupled TLSs which are immersed in either two IHBs or a CHB. We have
characterized the temperatures of the two coupled TLSs in eigenstate
and bare-state representations when the coupling between the two
TLSs is stronger and weaker than the TLS-bath couplings,
respectively. In the IHB case, we have found that, when the two IHBs
have the same temperatures, the two coupled TLSs could be
thermalized in eigenstate representation with the same temperature
as those of the heat baths. However, in the case where the two heat
baths have different temperatures, just when the energy detuning
between the two TLSs satisfies a special condition, the effective
four-level system formed by the two coupled TLSs can be thermalized
with an immediate temperature between those of the two heat baths.
In bare-state representation, we have found a counterintuitive
phenomenon that the temperature of the TLS connected with the
high-temperature heat bath is lower than that of the other TLS which
is connected with the low-temperature heat bath. In the CHB case,
the two TLSs in eigenstate representation could be thermalized with
the same temperature as that of the heat bath for nonresonant cases.
In bare state representation, we have found that the TLS with a
larger energy separation can be thermalized to a thermal equilibrium
at a lower temperature. We have also found a phenomenon of
anti-thermalization of the two TLSs in a common heat bath in the
resonant case. In addition, we have studied the steady-state
entanglement of the two TLSs in the IHB and CHB cases. It has been
found that there exist threshold temperatures for the steady-state
entanglement generation.

Finally, we present some discussions on the thermalization time over
which the thermalized systems evolve from their initial states to
steady states. Mathematically, the thermalization time for a system
should be infinite because the long-time limit
($\lim_{t\rightarrow\infty}$) is needed to make sure that these
density matrix elements evolve to their steady-state values. From
the viewpoint of physics, we might introduce some time scales to
describe a thermalization, as the half-life of an exponential decay.
However, for present systems, the evolutions of these density matrix
elements [i.e., the transient solution of Eq.~(\ref{OBEforihb})] are
not purely exponential functions. At the same time, these evolutions
depend on the initial conditions. Therefore, it is needed to
introduce the time scales under given initial conditions, other than
a universal time scale for a quantum thermalization.

\acknowledgments Jie-Qiao Liao is grateful to Professor C. P. Sun
for many helpful discussions. This work is supported in part by NSFC
under Grant No.~11075050, NFRPC under Grant No.~2007CB925204, and
PCSIRT under Grant No.~IRT0964.

\begin{widetext}
\appendix
\section{\label{appeindependentbath}Derivation of quantum master equation~(\ref{masterequation})
for the IHB case}

In this appendix, we give a detailed derivation of quantum master
equation~(\ref{masterequation}), which describes the evolution of
the two TLSs immersed in two IHBs. In the interaction picture with
respect to $H_{0}=H_{\textrm{TLSs}}+H^{(\textrm{IHB})}_{B}$, the
interacting Hamiltonian~(\ref{nondiacouplingindepH}) becomes
\begin{eqnarray}
H^{(\textrm{IHB})}_{I}(t)=[\tau_{13}B_{13}(t)+\tau_{24}B_{24}(t)]e^{i\varepsilon
_{1}t}+[\tau_{12}B_{12}(t)+\tau_{34}B_{34}(t)]e^{i\varepsilon_{2}t}
+h.c.,\label{HihbIP}
\end{eqnarray}
where $\tau_{ij}=|\lambda_{i}\rangle\langle\lambda_{j}|$ and we
introduce the noise operators
\begin{eqnarray}
B_{12}(t)&=&\sin(\theta/2)A(t)+\cos(\theta/2)B(t),\hspace{0.5 cm}
B_{13}(t)=\cos(\theta/2)A(t)-\sin(\theta/2)B(t),\nonumber\\
B_{24}(t)&=&\cos(\theta/2)A(t)+\sin(\theta/2)B(t), \hspace{0.5 cm}
B_{34}(t)=-\sin(\theta/2)A(t)+\cos(\theta/2)B(t),
\end{eqnarray}
with
$A(t)=\sum_{j}g_{1j}a_{j}e^{-i\omega _{aj}t}$ and
$B(t)=\sum_{k}g_{2k}b_{k}e^{-i\omega_{bk}t}$. Under the Born-Markov
approximation~\cite{Breuer}, the quantum master equation reads
\begin{eqnarray}\dot{\rho}_{S}=-\int_{0}^{\infty }dt'
\textrm{Tr}_{B}\left[H^{(\textrm{IHB})}_{I}(t),\left[H^{(\textrm{IHB})}_{I}(t-t')
,\rho _{S}(t) \otimes \rho _{B}\right]\right],
\end{eqnarray}
where $\textrm{Tr}_{B}$ stands for tracing over the degrees of
freedom of the baths. We assume that the two baths are in thermal
equilibrium state $\rho_{B}=\rho^{(a)}_{th}\otimes\rho^{(b)}_{th}$
with $\rho^{(a)}_{th}=Z^{-1}_{a}\exp(-\beta_{1}H^{(a)}_{B})$ and
$\rho^{(b)}_{th}=Z^{-1}_{b}\exp(-\beta_{2}H^{(b)}_{B})$, where
$Z_{a}=\textrm{Tr}_{B_{a}}[\exp(-\beta_{1}H^{(a)}_{B})]$ and
$Z_{b}=\textrm{Tr}_{B_{b}}[\exp(-\beta_{2}H^{(b)}_{B})]$ are the
partition functions of the two baths, respectively. The parameters
$\beta_{1}=1/T_{1}$ and $\beta_{2}=1/T_{2}$ are the inverse
temperatures of the baths for TLS$1$ and TLS$2$. Through making the
rotating wave approximation, we obtain
\begin{eqnarray}
\dot{\rho}_{S}&=&\sum_{(i,j)}\left[\tau _{ij}\rho _{S}\tau
_{ji}\int_{0}^{\infty }dt' e^{i(E_{\lambda_{i}}-E_{\lambda_{j}})t'
}\langle B_{ij}^{\dag }(-t') B_{ij}(0)\rangle-\tau _{jj}\rho
_{S}\int_{0}^{\infty }dt'
e^{-i(E_{\lambda_{i}}-E_{\lambda_{j}})t'}\langle B_{ij}^{\dag }(0)B_{ij}(-t')\rangle\right.\nonumber\\
&&+\left.\tau _{ji}\rho _{S}\tau _{ij}\int_{0}^{\infty }dt'
e^{-i(E_{\lambda_{i}}-E_{\lambda_{j}})t'}\langle
B_{ij}(-t')B^{\dag}_{ij}(0)\rangle-\tau _{ii}\rho
_{S}\int_{0}^{\infty }dt'
e^{i(E_{\lambda_{i}}-E_{\lambda_{j}})t' }\langle B_{ij}(0)B^{\dag }_{ij}(-t')\rangle\right]\nonumber\\
&&+\sum_{(ij,kl)}\left[\tau _{ij}\rho _{S}\tau _{kl}\int_{0}^{\infty
}dt' e^{i(E_{\lambda_{i}}-E_{\lambda_{j}})t' }\langle B_{lk}^{\dag
}(-t') B_{ij}(0)\rangle+\tau _{lk}\rho _{S}\tau
_{ji}\int_{0}^{\infty }dt' e^{i(E_{\lambda_{i}}-E_{\lambda_{j}})t'
}\langle B_{ij}^{\dag }(-t') B_{lk}(0)\rangle\right]
+h.c.,\label{mastereqforoff-diagonalid}
\end{eqnarray}
where the summation parameter $(i,j)$ in the first line of
Eq.~(\ref{mastereqforoff-diagonalid}) can take
$(i,j)=(1,2),(1,3),(2,3)$, and $(2,4)$, and the summation parameter
$(ij,kl)$ in the third line of Eq.~(\ref{mastereqforoff-diagonalid})
can take $(ij,kl)=(12,43),(13,42),(31,24)$, and $(43,12)$. Here the
bath correlation functions are defined by $\langle
X(t)Y(t')\rangle=\textrm{Tr}_{B}[X(t)Y(t')\rho_{B}]$, and we use the
property $\langle X(t)Y(t')\rangle=\langle
X(t-t')Y(0)\rangle=\langle X(0)Y(t'-t)\rangle$ of the correlation
functions. To derive the quantum master equation, we need to
calculate the one-side Fourier transform of the correlation
functions in Eq.~(\ref{mastereqforoff-diagonalid}). For simplicity,
in the following we only keep the real parts of the one-side Fourier
transforms of the correlation functions and neglect their imaginary
parts since the imaginary parts only contribute to the Lamb shifts,
which are neglected in this work. The real parts of the one-side
Fourier transform of the correlation functions can be obtained as
follows:
\begin{eqnarray}
&&\textrm{Re}\left[\int_{0}^{\infty }dt' e^{i\varepsilon _{1}t'
}\langle B_{24}(0)B^{\dag}_{24}(-t')\rangle\right]=
\textrm{Re}\left[\int_{0}^{\infty }dt' e^{i\varepsilon _{1}t'
}\langle
B_{13}(0)B^{\dag}_{13}(-t')\rangle\right]=\Gamma_{1},\nonumber\\
&&\textrm{Re}\left[\int_{0}^{\infty }dt' e^{-i\varepsilon _{1}t'
}\langle
B_{24}^{\dag}(0)B_{24}(-t')\rangle\right]=\textrm{Re}\left[\int_{0}^{\infty
}dt' e^{-i\varepsilon _{1}t'
}\langle B_{13}^{\dag}(0)B_{13}(-t')\rangle\right]=\Gamma_{2},\nonumber\\
&&\textrm{Re}\left[\int_{0}^{\infty }dt' e^{i\varepsilon _{2}t'
}\langle B_{12}(0)B^{\dag}_{12}(-t')\rangle\right]=
\textrm{Re}\left[\int_{0}^{\infty }dt' e^{i\varepsilon _{2}t'
}\langle
B_{34}(0)B^{\dag}_{34}(-t')\rangle\right]=\Gamma_{3},\nonumber\\
&&\textrm{Re}\left[\int_{0}^{\infty }dt' e^{-i\varepsilon _{2}t'
}\langle B_{12}^{\dag}(0)B_{12}(-t')\rangle\right]=
\textrm{Re}\left[\int_{0}^{\infty }dt' e^{-i\varepsilon _{2}t'
}\langle B_{34}^{\dag }(0)B_{34}(-t')\rangle\right]=\Gamma_{4},
\end{eqnarray}
and
\begin{eqnarray}
&&\textrm{Re}\left[\int_{0}^{\infty }dt' e^{-i\varepsilon_{1}t'
}\langle B_{24}(-t')B^{\dag}_{13}(0)\rangle\right]
=\textrm{Re}\left[\int_{0}^{\infty }dt' e^{-i\varepsilon _{1}t'
}\langle B_{13}(-t')B^{\dag}_{24}(0)\rangle\right]=\Lambda_{1},\nonumber\\
&&\textrm{Re}\left[\int_{0}^{\infty }dt' e^{-i\varepsilon _{2}t'
}\langle B_{12}(-t')B^{\dag}_{34}(0)\rangle\right]
=\textrm{Re}\left[\int_{0}^{\infty }dt' e^{-i\varepsilon _{2}t'
}\langle B_{34}(-t')B^{\dag}_{12}(0)\rangle\right]=\Lambda_{2},\nonumber\\
&&\textrm{Re}\left[\int_{0}^{\infty }dt' e^{i\varepsilon _{1}t'
}\langle B^{\dag}_{24}(-t')B_{13}(0)\rangle\right]
=\textrm{Re}\left[\int_{0}^{\infty }dt' e^{i\varepsilon _{1}t'
}\langle
B^{\dag}_{13}(-t')B_{24}(0)\rangle\right]=\Lambda_{3},\nonumber\\
&&\textrm{Re}\left[\int_{0}^{\infty }dt' e^{i\varepsilon
_{2}t'}\langle B^{\dag}_{12}(-t')B_{34}(0)\rangle\right]
=\textrm{Re}\left[\int_{0}^{\infty }dt' e^{i\varepsilon _{2}t'
}\langle B^{\dag}_{34}(-t')B_{12}(0)\rangle\right]=\Lambda_{4},
\end{eqnarray}
where the parameters $\Gamma_{i}$ and $\Lambda_{i}$ have been
defined in Eq.~(\ref{defofGmmaandLamb}). Based on the above one-side
Fourier transforms of these correlation functions, we can obtain
those for other correlation functions. By substituting them into
quantum master equation~(\ref{mastereqforoff-diagonalid}) and
returning to the Schr\"{o}dinger picture, we can obtain quantum
master equation~(\ref{masterequation}).

In the following we give an example for calculation of the one-side
Fourier transform of correlation function,
\begin{eqnarray}
\textrm{Re}\left[\int_{0}^{\infty }dt' e^{-i\varepsilon _{1}t'
}\langle B_{24}^{\dag}(0)B_{24}(-t')\rangle\right]
&=&\cos^{2}(\theta/2)\textrm{Re}\left[\int_{0}^{\infty }dt'
e^{-i\varepsilon _{1}t' }\langle
A^{\dag}(0)A(-t')\rangle\right]\nonumber\\
&&+\sin^{2}(\theta/2)\textrm{Re}\left[\int_{0}^{\infty }dt'
e^{-i\varepsilon _{1}t' }\langle B^{\dag}(0)B(-t')\rangle\right],
\end{eqnarray}
which is based on the fact that there is no correlation between the
two IHBs. We can calculate
\begin{eqnarray}
\textrm{Re}\left[\int_{0}^{\infty }dt' e^{-i\varepsilon _{1}t'
}\langle A^{\dag}(0)A(-t')\rangle\right]
=\sum_{j}g^{2}_{1j}\bar{n}_{a}(\omega_{aj})\pi\delta(\omega_{aj}-\varepsilon_{1})
=\pi\varrho(\varepsilon_{a})g^{2}_{1}(\varepsilon_{1})\bar{n}_{a}(\varepsilon_{1})
=\gamma_{a}(\varepsilon_{1})\bar{n}_{a}(\varepsilon_{1}),
\end{eqnarray}
where we introduce the rate
$\gamma_{a}(\varepsilon_{1})=\pi\varrho_{a}(\varepsilon_{1})g^{2}_{1}(\varepsilon_{1})$
and the average thermal excitation
$\bar{n}_{a}(\varepsilon_{1})=1/[\exp(\varepsilon_{1}/T_{1})-1]$.
Note that here we have also used the formula
\begin{eqnarray}
\int_{0}^{\infty }dt' e^{\pm i\omega t' }=\pi\delta(\omega)\pm
i\mathbf{P}\frac{1}{\omega },
\end{eqnarray}
where the sign ``$\mathbf{P}$" stands for the principal value
integral. Similarly, we can obtain
$\textrm{Re}\left[\int_{0}^{\infty }dt' e^{-i\varepsilon _{1}t'
}\langle B^{\dag}(0)B(-t')\rangle\right]
=\gamma_{b}(\varepsilon_{1})\bar{n}_{b}(\varepsilon_{1})$ with
$\gamma_{b}(\varepsilon_{1})=\pi\varrho_{b}(\varepsilon_{1})g^{2}_{2}(\varepsilon_{1})$.
Therefore, we have
\begin{eqnarray}
\textrm{Re}\left[\int_{0}^{\infty }dt' e^{-i\varepsilon _{1}t'
}\langle B_{24}^{\dag}(0)B_{24}(-t')\rangle\right]=\cos^{2}( \theta
/2)\gamma_{a}(\varepsilon_{1})\bar{n}_{a}(\varepsilon _{1})+\sin
^{2}(\theta/2)\gamma_{b}(\varepsilon
_{1})\bar{n}_{b}(\varepsilon_{1})=\Gamma_{2}.
\end{eqnarray}
With the same method, the one-side Fourier transform for other
correlation functions can also be obtained.

\section{\label{appecommontbath}Derivation of the rates in Eq.~(\ref{ratesforcommonbath}) for the CHB case}

In the CHB case, the interaction Hamiltonian has the same form as
Eq.~(\ref{HihbIP}). But now the noise operators become
\begin{eqnarray}
B_{12}(t)&=&\sin(\theta/2) A_{1}(t)
+\cos(\theta/2)A_{2}(t),\hspace{0.5 cm}
B_{13}(t)=\cos(\theta/2) A_{1}(t)-\sin(\theta/2)A_{2}(t),\nonumber\\
B_{24}(t)&=&\cos(\theta/2) A_{1}(t)+\sin(\theta/2)
A_{2}(t),\hspace{0.5 cm}
B_{34}(t)=-\sin(\theta/2)A_{1}(t)+\cos(\theta/2) A_{2}(t),
\end{eqnarray}
with $A_{1}(t)=\sum_{j}g_{1j}a_{j}e^{-i\omega_{j}t}$ and $
A_{2}(t)=\sum_{j}g_{2j}a_{j}e^{-i\omega_{j}t}$. The quantum master
equation for the CHB case has the same form as
Eq.~(\ref{mastereqforoff-diagonalid}), but now the real parts of the
one-side Fourier transform for the correlation functions become
\begin{eqnarray}
\textrm{Re}\left[\int_{0}^{\infty}e^{-i\varepsilon _{2}t'}\langle
B_{12}(-t')B_{12}^{\dag }(0)\rangle
dt'\right]&=&\Gamma_{12},\hspace{0.5 cm}
\textrm{Re}\left[\int_{0}^{\infty }e^{-i\varepsilon _{1}t' }\langle
B_{13}(-t')B_{13}^{\dag }(0)\rangle dt'
\right]=\Gamma_{13},\nonumber\\
\textrm{Re}\left[ \int_{0}^{\infty }e^{-i\varepsilon _{1}t' }\langle
B_{24}(-t')B_{24}^{\dag }(0)\rangle dt' \right]
&=&\Gamma_{24},\hspace{0.5 cm} \textrm{Re}\left[ \int_{0}^{\infty
}e^{-i\varepsilon _{2}t' }\langle B_{34}( -t')B_{34}^{\dag
}(0)\rangle dt' \right] =\Gamma_{34},\nonumber\\
\textrm{Re}\left[\int_{0}^{\infty }e^{i\varepsilon _{2}t' }\langle
B_{12}^{\dag}(-t')B_{12}(0)\rangle dt'
\right]&=&\Gamma_{21},\hspace{0.5 cm}
\textrm{Re}\left[\int_{0}^{\infty }e^{i\varepsilon _{1}t' }\langle
B_{13}^{\dag}(-t')B_{13}(0)\rangle dt'
\right]=\Gamma_{31},\nonumber\\
\textrm{Re}\left[\int_{0}^{\infty }e^{i\varepsilon _{1}t' }\langle
B_{24}^{\dag}(-t')B_{24}(0)\rangle dt'
\right]&=&\Gamma_{42},\hspace{0.5 cm}
\textrm{Re}\left[\int_{0}^{\infty }e^{i\varepsilon _{2}t' }\langle
B_{34}^{\dag}(-t')B_{34}(0)\rangle dt' \right]=\Gamma_{43},
\end{eqnarray}
and
\begin{eqnarray}
\textrm{Re}\left[ \int_{0}^{\infty }e^{-i\varepsilon _{1}t' }\langle
B_{24}\left(-t' \right)B_{13}^{\dag }\left(0\right)\rangle dt'
\right]&=&\textrm{Re}\left[ \int_{0}^{\infty }e^{-i\varepsilon
_{1}t' }\langle B_{13}\left( -t' \right)B_{24}^{\dag
}\left(0\right)\rangle dt' \right]
=\Lambda_{1},\nonumber\\
\textrm{Re}\left[ \int_{0}^{\infty }e^{-i\varepsilon _{2}t' }\langle
B_{12}\left(-t' \right)B_{34}^{\dag }\left(0\right)\rangle dt'
\right]&=&\textrm{Re}\left[ \int_{0}^{\infty }e^{-i\varepsilon
_{2}t' }\langle B_{34}\left(-t' \right)B_{12}^{\dag
}\left(0\right)\rangle dt'
\right]=\Lambda_{2},\nonumber\\
\textrm{Re}\left[ \int_{0}^{\infty }e^{i\varepsilon _{1}t' }\langle
B_{24}^{\dag }(-t')B_{13}(0)\rangle dt' \right]&=&\textrm{Re}\left[
\int_{0}^{\infty }e^{i\varepsilon _{1}t' }\langle B_{13}^{\dag
}(-t')B_{24}(0)\rangle dt' \right]
=\Lambda_{3},\nonumber\\
\textrm{Re}\left[ \int_{0}^{\infty }e^{i\varepsilon _{2}t' }\langle
B_{34}^{\dag }(-t') B_{12}(t)\rangle dt' \right]
&=&\textrm{Re}\left[ \int_{0}^{\infty }e^{i\varepsilon _{2}t'
}\langle B_{12}^{\dag}(-t') B_{34}(0) \rangle dt' \right]
=\Lambda_{4},
\end{eqnarray}
where the parameters $\Gamma_{ij}$ and $\Lambda_{i}$ have been
defined in Eq.~(\ref{ratesforcommonbath}). The one-side Fourier
transform of other correlation functions can be obtained in terms of
the above results. Below, we give an example for calculation of the
one-side Fourier transform of correlation functions,
\begin{eqnarray}
&&\textrm{Re}\left[\int_{0}^{\infty }e^{i\varepsilon _{2}t' }\langle
B_{12}^{\dag}(-t')B_{12}(0)\rangle dt' \right]\nonumber\\
&=&\sin
^{2}\left( \theta /2\right)\textrm{Re}\left[\int_{0}^{\infty
}e^{i\varepsilon _{2}t' }\langle A_{1}^{\dag}(-t')A_{1}(0)\rangle
dt' \right]+\cos ^{2}\left( \theta
/2\right)\textrm{Re}\left[\int_{0}^{\infty }e^{i\varepsilon _{2}t'
}\langle A_{2}^{\dag}(-t')A_{2}(0)\rangle dt'
\right]\nonumber\\&&+\frac{1}{2}\sin \theta
\left\{\textrm{Re}\left[\int_{0}^{\infty }e^{i\varepsilon _{2}t'
}\langle A_{1}^{\dag}(-t')A_{2}(0)\rangle dt'
\right]+\textrm{Re}\left[\int_{0}^{\infty }e^{i\varepsilon _{2}t'
}\langle A_{2}^{\dag}(-t')A_{1}(0)\rangle dt' \right]\right\}.
\end{eqnarray}
We calculate
\begin{eqnarray}
\textrm{Re}\left[\int_{0}^{\infty }dt'
e^{i\varepsilon_{2}t' }\langle
A_{1}^{\dag}(-t')A_{1}(0)\rangle\right]
=\sum_{j}g^{2}_{1j}\bar{n}(\omega_{j})\pi\delta(\omega_{j}-\varepsilon_{2})
=\pi\varrho(\varepsilon_{2})g^{2}_{1}(\varepsilon_{2})\bar{n}(\varepsilon_{2})
=\gamma_{1}(\varepsilon_{2})\bar{n}(\varepsilon_{2}),
\end{eqnarray}
where we introduce the rate
$\gamma_{1}(\varepsilon_{2})=\pi\varrho(\varepsilon_{2})g^{2}_{1}(\varepsilon_{2})$
and the average thermal excitation
$\bar{n}(\varepsilon_{2})=1/[\exp(\varepsilon_{2}/T)-1]$. Using the
same method, we can obtain
\begin{eqnarray}
\textrm{Re}\left[\int_{0}^{\infty }dt' e^{i\varepsilon_{2}t'
}\langle A_{1}^{\dag}(-t')A_{2}(0)\rangle\right]
=\sum_{j}g_{1j}g_{2j}\bar{n}(\omega_{j})\pi\delta(\omega_{j}-\varepsilon_{2})
=\pi\varrho(\varepsilon_{2})g_{1}(\varepsilon_{2})g_{2}(\varepsilon_{2})\bar{n}(\varepsilon_{2})
=\gamma_{12}(\varepsilon_{2})\bar{n}(\varepsilon_{2}),
\end{eqnarray}
where the rate
$\gamma_{12}(\varepsilon_{2})=\pi\varrho(\varepsilon_{2})g_{1}(\varepsilon_{2})g_{2}(\varepsilon_{2})=\sqrt{\gamma_{1}(\varepsilon_{2})\gamma_{2}(\varepsilon_{2})}$.
Similarly, we have $\textrm{Re}\left[\int_{0}^{\infty }dt'
e^{i\varepsilon_{2}t'}\langle
A_{2}^{\dag}(-t')A_{2}(0)\rangle\right]=\gamma_{2}(\varepsilon_{2})\bar{n}(\varepsilon_{2})$
and $\textrm{Re}\left[\int_{0}^{\infty }dt' e^{i\varepsilon_{2}t'
}\langle
A_{2}^{\dag}(-t')A_{1}(0)\rangle\right]=\gamma_{12}(\varepsilon_{2})\bar{n}(\varepsilon_{2})$.
Therefore, we obtain
\begin{eqnarray}
\textrm{Re}\left[\int_{0}^{\infty }e^{i\varepsilon _{2}t' }\langle
B_{12}^{\dag}(-t')B_{12}(0)\rangle dt'
\right]=\left[\sin(\theta/2)\sqrt{\gamma_{1}(\varepsilon_{2})}
+\cos(\theta/2)\sqrt{\gamma_{2}(\varepsilon_{2})}
\right]^{2}\bar{n}(\varepsilon_{2})=\Gamma_{21}.
\end{eqnarray}
The one-side Fourier transform for other correlation functions can
also be obtained with the same method.
\end{widetext}

\end{document}